# Application of Probabilistic Graphical Models in Forecasting Crude Oil Price

by

Danish A. Alvi

A dissertation submitted in partial satisfaction of the

requirements for the degree of

**Bachelors of Science**

in

Computer Science

Department of Computer Science

**University College London**

Spring 2018

# Abstract

The dissertation investigates the application of Probabilistic Graphical Models (PGMs) in forecasting the price of Crude Oil.

This research is important because crude oil plays a very pivotal role in the global economy hence is a very critical macroeconomic indicator of the industrial growth. Given the vast amount of macroeconomic factors affecting the price of crude oil such as supply of oil from OPEC countries, demand of oil from OECD countries, geopolitical and geoeconomic changes among many other variables - probabilistic graphical models (PGMs) allow us to understand by learning the graphical structure. This dissertation proposes condensing data numerous Crude Oil factors into a graphical model in the attempt of creating a accurate forecast of the price of crude oil.

The research project experiments with using different libraries in Python in order to construct models of the crude oil market. The experiments in this thesis investigate three main challenges commonly presented while trading oil in the financial markets. The first challenge it investigates is the process of learning the structure of the oil markets; thus allowing crude oil traders to understand the different physical market factors and macroeconomic indicators affecting crude oil markets and how they are *causally* related. The second challenge it solves is the exploration and exploitation of the available data and the learnt structure in predicting the behaviour of the oil markets. The third challenge it investigates is how to validate the performance and reliability of the constructed model in order for it to be deployed in the financial markets.

A design and implementation of a probabilistic framework for forecasting the price of crude oil is also presented as part of the research.

## Contributions to Science

The primary contribution of this dissertation to the Quantitative Investment Strategy research are the robust Bayesian-based models for the oil market to handle challenges in forecasting the spot price of crude oil in real world trading environment for portfolio managers. Computational Finance and Bayesian Methodologies are proposed to:

1. Identify the different macroeconomic, microeconomic and geopolitical variables that affect the price of oil globally.

2. Determine the causal relationship between these different variables and creating a probabilistic graphical model to represent the oil market dynamics.

3. Evaluating and testing the performance of this model by stress testing: simulating situations of economic distress in order to assess the reliability of the model.

This thesis contributes to existing literature in a number of ways: First, it allows us to forecast the price of crude oil using a global-macro strategy, using macroeconomic data and physical market factors of the oil markets. Secondly, it proposes a method of learning the structure of the oil markets and therefore allows us to crunch a large amount of data. Lastly, it allows us to understand how the oil markets work and therefore assists economists and energy policy makers to determine which factors to control and what policies to draft in an attempt to prevent the economy from going in a recession and to control climate change by reducing dependency on fossil fuels.

# Contents

# Chapter 1

# Introduction

This chapter introduces the general context of the dissertation. The motivation of presenting this research to accurately predict the behaviour of the energy markets using a global macro strategy hence assist portfolio managers and energy traders in making better investment decisions. It is our objective to use probabilistic graphical models to construct a model of the oil markets and carry out inferences on them to forecast multi-quarter behaviour of the oil markets, and therefore we propose using Belief Networks and Hidden Markov Models. We also propose using U.S. Federal Government open-data facilities to retrieve relevant macroeconomic and physical market factors.

## 1.1   Motivation

The application of Machine Learning in Quantitative and Computational Finance has become increasingly popular in the recent years. An increasing number of hedge funds such as Man Group, Two Sigma Investments, Winton Capital, Renaissance Technologies and a number of investment banks such as Goldman Sachs Asset Management (GSAM) and Bank of America Merrill Lynch have been applying Machine Learning and Artificial Intelligence in their trades. In 2015, Artificial Intelligence was contributing roughly half the profits in one of MANs biggest funds, the AHL Dimension Programme. Two Sigma Investments, a New York based hedge fund, is entirely basing its revenues from strategies based on Machine Learning.

Hedge funds employing a global macro strategy observe minute changes in the macroeconomic behaviour, in which the price of crude oil is a vital key player. Crude oil plays an important role in the macroeconomic stability given not only its utilisation in conventional fuels but also its utilisation in creating infrastructure, such as the use of bitumen in laying roads. Therefore, in the long-term, crude oil may heavily influence the rate of economic growth of countries, especially those relying on oil imports and hence have a heavy influence on the performance of the global financial markets. There is widespread agreement that unexpected fluctuations in the real price of crude oil are detrimental to the welfare of both oil-importing and oil-exporting economies. Reliable forecasts of the price of oil are of interest for a wide range of applications. For example, global macro hedge-funds view forecast the price of oil as one of the key variables in generating macroeconomic projections and assessing the general macroeconomic atmosphere of the global financial markets. The price of oil plays an important role for policy makers in predicting recessions. For example, Hamilton (2009), building on the analysis in Edelstein and Kilian (2009), provides evidence that the 2008 Financial Crisis was preceded by an economic slowdown in the automobile industry and a consequent depreciation of consumer sentiment.

With the exponentially increasing amount of datasets and the increasing amount of computational power and storage being available in the form of distributed and cloud computing, the cost of storing, crunching, and analysing Big Data has been decreasing and therefore it has become even more feasible to use Bayesian Analysis for financial forecasting. Therefore, many hedge funds, such as Two Sigma investments, are actively harnessing the benefits of distributed Cloud Computing, Machine Learning and Artificial Intelligence. Although it is true that the performance of the model tends to increase with more data, due to Friedrich Hayeks Dispersed Knowledge notion, the performance of a Bayesian model might increase until a certain number of datasets have been incorporated, after which the model will begin showing traits of over-fitness, which is not desirable.

The first problem is that the existing models for analysing the spot price for crude oil such as GARCH are based on the mere assumption that the error variance of the spot price of crude Oil forecast follows the ARMA model, and therefore these models certainly do not incorporate the macroeconomic, microeconomic and geopolitical events which play an important role in determining the spot price for crude oil. The research proposes using Bayesian derived views which be compared with the GARCH based views to determine which scheme is better.

The second problem is that given the vast amount of datasets available on the Energy Information administration's website, it is not only very time consuming task for a quantitative analyst to relate all the data using expert knowledge, but it is also a very error-prone task and the Bayesian Probabilistic Graphical Model would always be changing in the face of new and changing market dynamics. The research proposes a scheme which would allow the system to learn the Bayesian Network in an attempt of causally relating all datasets without the presence of an expert.

The third problem is regarding how the futures price of crude oil is related to forecast of the spot price and how it could be used to analyse the volatility of oil. Given that the futures price reflects the price buyers are willing to pay for oil on a delivery date set at some point in the future, it is an indicator of the sentiment of the buyers and hence is a very useful tool that could be used for forecasting the price of oil. We would be using Machine Learning Regression analytical techniques in order to determine the relation between futures price and spot price and will incorporate it in our Bayesian Model.

The main motivation of this research is to create a Bayesian Model of the oil markets using Probabilistic Graphical Models in an attempt improve the existing quantitative models for the oil markets.

## 1.2   Objective of Research

The objective of the research is to propose a model for accurately forecasting the price of crude oil by representing structural and macroeconomic changes in the oil market by using Probabilistic Graphical Models. The main hypothesis of the research is:

> By using Probabilistic Graphical Models and applying various Machine Learning and Statistical techniques, the accuracy of existing quantitative models forecasting the spot price of oil can be improved by taking in account many different macroeconomic and geopolitical variables in consideration when making a prediction system.

To validate this hypothesis, there are **four** primary tasks for the research:

## 1.2.1 Understanding the structure of the energy markets on a macroeconomic level

The first task is to understand the macroeconomic structure of the oil markets to have an idea of the factors affect the price of crude oil. There are a number of macroeconomic factors [42] which affect the energy markets such as production, consumption, policy, geopolitics, GDP Growth, Consumer Price Index (CPI) and Industrial Production Index (IPI) which determine the direction of the energy markets.

## 1.2.2 Exploring government open-data facilities by extracting relevant datasets

The second task is to fetch the relevant datasets from governmental open-data sources which are most relevant to understanding the macroeconomic situation of the energy markets. Datasets describing these macroeconomic indicators are easily available on governmental open-data facilities and can be easily retrieved by API calls from Python.

## 1.2.3 Constructing a Bayesian Network of Energy Markets using the macroeconomic indicators.

The third task is to construct a Bayesian Belief network based on the datasets collected from the open-data facilities. Constructing the Bayesian Network can be achieved in a number of ways, such as by either using expert knowledge or using network learning methods such as Hill Climb Search [50], or perhaps a combination of both. We would be using libraries in Python which would allow us to easily implement Probablistic Graphic Models such as Bayesian Networks and carry out the necessary inferential operations to make forecasts.

### 1.2.4 Testing the outcome of the Bayesian Model based on historical performance and results.

The fourth task is testing our constructed Bayesian Network by historical backtesting on oil price data. Stress testing [56] plays an important role in the backtesting process as it allows the portfolio managers to understand the behaviour of the Bayesian model in volatile markets. We would initially be using a basic model of comparing the predicted inferences from test data with the actual outcome of the predicted variable and outputing it as a percentage, and later would be using some Python libraries for backtesting and even live testing our data.

## 1.3 Data sources used in this research

The United States Federal Government [61] has a number of open-data services which provide macroeconomic time-series data about the production, consumption, energy policy, and financial atmosphere of the industry. We would be primarily using two data sources in our research, mainly the Energy Information Administration for obtaining data regarding oil, and the Federal Reserve Bank Economic Data for datasets concerning macroeconomic factors.

### 1.3.1 Federal Reserve Economic Data, St. Louis

The Federal Reserve Economic Data, St. Louis (FRED) is a database maintained by the research division of Federal Reserve Bank of St. Louis which contains a number of economic time-series data. The FRED has a number of macroeconomic indicators such as the Effective Funds Rate and Consumer Price Index (CPI), which are popularly used in quantitative models. We would be using the FRED's data for retrieving data sets concerning macroeconomic indicators which directly affect the price of oil such as the Consumer Price Index (CPI), global interest rates, and global industrial production.

### 1.3.2 Energy Information Administration

The Energy Information Administration (EIA) is the primary organisation of the United States Government which is responsible for collecting, analysing, and disseminating energy information to promote policy making, efficient markets and the public understanding of the relationship of energy with the economy and the environment. We would be using the EIAs data for not only retrieving data sets regarding the current and future supply, demand of oil in the global markets, but also expert knowledge for constructing the Probabilistic Graphical Models relating these datasets.

# Chapter 2

# Background & Literature Review

This chapter introduces the general context of the application of *Probabalistic Graphical Models* in forecasting. The motivation of presenting this research is to *accurately predict the behaviour of the energy markets* hence assist portfolio managers and energy traders in making better investment decisions. The application of Probabilistic Graphical Models in making predictions has been discussed. Moreover, the *macroeconomic structure of the oil markets* and the data sources from where the macroeconomic data is retrieved has also been discussed. Furthermore, the Python libraries being used for Bayesian Programming are also discussed.

## 2.1 Probabilistic Graphical Models

Probablistic Graphical Models (PGMs) are a *graphical* illustration of a *joint probability distribution* which exploits dependencies between different *random variables*.

PGMs allow us to describe the affect of one variable on the other and conversely, the independence of variables. The intersection of *Graph Theory* and *Probability Theory* allows us to construct better models that are not only better specified but also computationally more efficient. PGMs are essentially the outcome of giving a *Graphical Structure* to variables of a probabilistic model and this process captures the relationships between variables and their uncertainties, conditional dependencies and independence.

There are two groups of Graphical Models: *belief networks* and *Markov networks*. The project would be applying both these Graphical Models as they are highly applicable in the construction of stochastic and probabilistic models of the energy markets. In this section, we would we introducing both these graphical models through definitions and examples and their applicability in making predictions in a general context.

### 2.1.1 Belief Networks

BELIEF NETWORKS (also known as *Bayesian Networks* or *Bayesian Belief Networks*) are a way of illustrating the independence assumptions made in a probabilistic distribution. A belief network is primarily used to represent *causal* relationships between random variables.

Causality is can be illustrated via trivial example[24]. Suppose we attempt to switch on a computer, but we make an *observation* that the computer does not switch on. We would like to know what could be the possible reasons of the failure. In a simplified model, we assume there are only **two** causes: *electricity failure*, *E*, and *computer malfunction*, *M* to the observation of *computer failure*, *C*. Below is an representation the *graphical structure* of the causal model we have assumed.

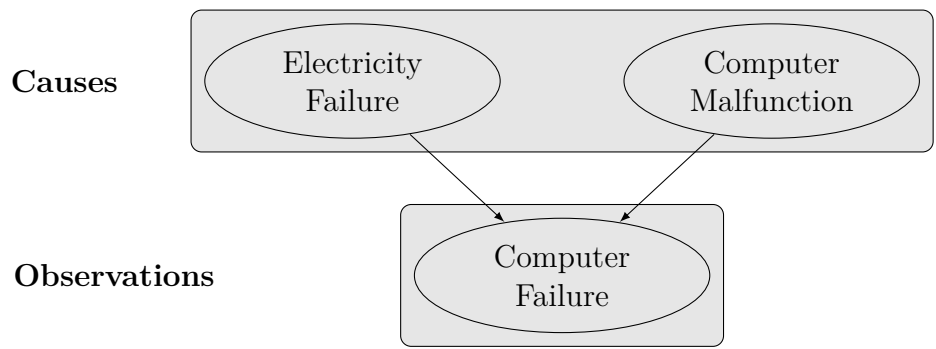

Figure 2.1: A graphical illustration of the structure of causal model.

**Bayes Theorem** allows us to calculate the conditional probability distribution of the *unobserved causes* given the *observed evidence*.

---

**Theorem 1 (Bayes Theorem)** Given an unobserved cause $H$ and observed evidence $E$, the relationship between the posterior probability [1], likelihood [2], priori probability [3], and marginal probability [4] is given by:

$$P(H \mid E) = P(E \mid H) \cdot \frac{P(H)}{P(H)}$$

---

Refering back to our example of from Cowell [24], we can see the the *causes* are 'electricity failure' and 'computer malfunction', whereas the *observations* are 'computer failure'. Assume that '$E$ occurs with probability $P(E = true) = 0.1$, and $M$, occurs with probability $P(M = true) = 0.2$. Also assume that the property $P(E, M) = P(E) \cdot P(M)$ holds (i.e. $E$ and $M$ are independent). Also assume (for the sake of *simplicity*) that the if there is no problem with electricity and the computer has no malfunction, then $P(C = true \mid E = false, M = false)$. Lastly, we assume that if there is an electricity malfunction, the computer will not start regardless the if there is a computer malfunction, i.e. $P(C = true \mid E = true, M = false) = 1$ and $P(C = true \mid E = true, M = true) = 1$. Given these assumptions, we can calculate the the *priori* probability $P(C = true)$ using *marginalisation* [5].

$$
\begin{aligned}
P[C = true] &= \sum_{E,M} P[C = true, E, M] \\
&= \sum_{E,M} P[C = true \mid E, M] \cdot P[E] \cdot P[M] = \mathbf{0.19}
\end{aligned}
$$

Now, using this result, we can draw conculsions on the *unobserved causes* given the *observed evidence*. For example, if we try to switch on the computer but it does not start, we can calculate the probability distribution of the *unobserved causes* **given** the *observed evidence*.

---

[1] *Posterior* probability is the probability of the *unobserved cause after* knowing the *observed evidence*.

[2] *Likelihood* is the probability of the *unobserved evidence* given the *observed cause*.

[3] *Priori* probability is the probability of the *unobserved cause before* knowing the *observed evidence*.

[4] *Marginal* probability is the *total* probability of the *observed evidence*.

For calculating the probability that there has been an electricity failure (*unobserved cause*) **given** the computer is not switching on,

$$P[E = true \mid C = true] = \sum_M P[E = true, M \mid C = true] \qquad \textbf{Marginalisation}$$

$$= \sum_M \frac{P[C = true \mid E = true, M] \cdot P[E = true] \cdot P[M]}{P[C = true]} \quad \textbf{Bayes' Theorem}$$

$$= \frac{1 \cdot 0.1 \cdot 0.2}{0.19} + \frac{1 \cdot 0.1 \cdot 0.8}{0.19} = \frac{10}{19} \approx \textbf{0.53}$$

For calculating the probability that there has been an computer malfunction (*unobserved cause*) **given** the computer is not switching on,

$$P[M = true \mid C = true] = \sum_E P[E, M = true \mid C = true] \qquad \textbf{Marginalisation}$$

$$= \sum_E \frac{P[C = true \mid E, M = true] \cdot P[E] \cdot P[M = true]}{P[C = true]} \quad \textbf{Bayes' Theorem}$$

$$= \frac{1 \cdot 0.1 \cdot 0.2}{0.19} + \frac{0.5 \cdot 0.9 \cdot 0.2}{0.19} = \frac{11}{19} \approx \textbf{0.58}$$

**Belief Networks** heavily rely on *Bayes' Theorem* for calculating *joint probability distributions* of variables using *conditional probabilities* obtained from *Bayes' Theorem* .

---

**Definition 1 (Belief Network )** A Belief Network is a *directed acyclic graph* $G = \langle V, E \rangle$ where every vertex $v \in V$ is associated with a random variable $X_v$, and every edge $(u, v) \in E$ represents a direct dependence from a random variable $X_u$ to a random variable $X_v$ [31]. The probability distribution function is of the form[5]:

$$p(x_1, \cdots, x_D) = \prod_{i=1}^{D} p(x_i \mid \mathrm{pa}(x_i))$$

where $\mathrm{pa}(x_i)$ represents the parental variables of $x_i$, and the $i^{th}$ vertex in the graph corresponding with the factor $p(x_i \mid \mathrm{pa}(x_i))$ [5]. In a belief Network, each node $v \in V$ corresponds to a *Conditional Probability Table*, $\mathsf{CPD}(v)$, which denotes the probability

---

distribution of $X_v$, conditioned over the values of the random variables associated with the direct dependencies $D(v) = \{u \mid (u, v) \in E\}$.

In order to explain the concept of belief networks better, the most common example that is given is that of the "Rain-Sprinkler-Wet Grass" model, which is popularly described in literature in Statistics and Machine Learning [37].

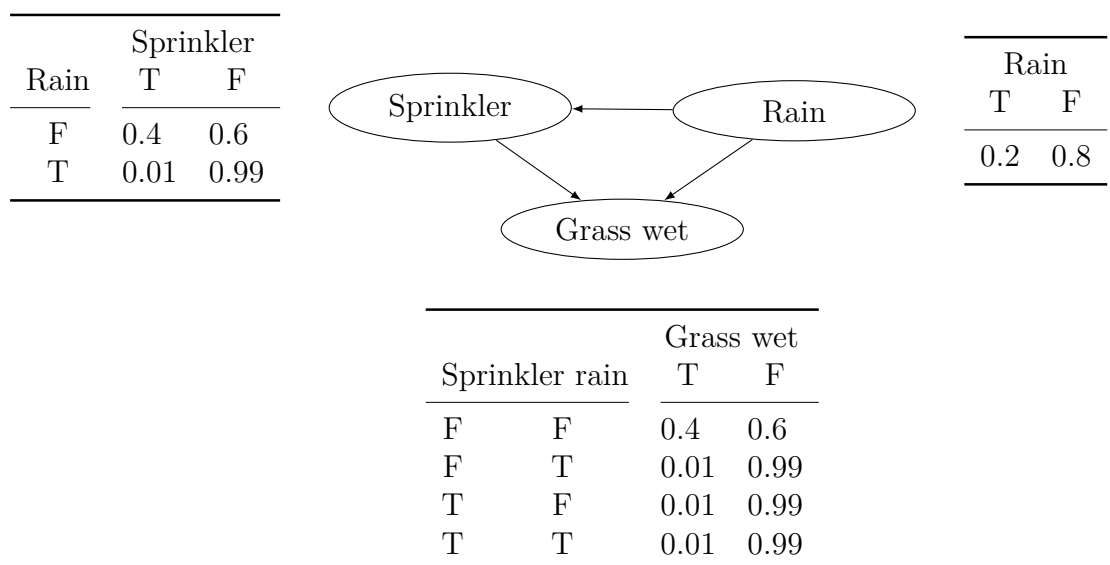

Figure 2.2: The 'Rain-Sprinkler-Wet Grass' belief network example.

In this model, we can use the *observed evidence* of the grass being wet to calculate the probabilities of the *unobserved causes*. For example, we can calculate the probability of the sprinkler being on (*evidence*) given the grass is wet (*observation*).

$$P[R = T \mid G = T] = \frac{P[G = T, R = T]}{P[G = T]} \qquad \textbf{Bayes' Theorem}$$

$$= \frac{\sum_S P[G = T, S, R = T]}{\sum_{S,R} P[G = T, S, R]} \qquad \textbf{Marginalisation}$$

---

[1]In the case of a random variable having no parents, $p(x_i \mid \mathrm{pa}(x_i)) = p(x_i)$.

$$\sum_S P[G=T, S, R=T] = P[G=T, S=T, R=T] \ +$$

$$= P[G=T, S=F, R=T]$$

$$= P[G=T \mid S=T, R=T] \cdot P[S=T \mid R=T] \cdot P[R=T] \ +$$

$$P[G=T \mid S=F, R=T] \cdot P[S=F \mid R=T] \cdot P[R=T] \qquad \textbf{Joint P.D.F}$$

$$= 0.01 \cdot 0.01 \cdot 0.2 + 0.01 \cdot 0.99 \cdot 0.2 = \textbf{0.002}$$

$$\sum_S P[G=T, S, R] = P[G=T, S=T, R=T] \ +$$

$$P[G=T, S=T, R=F] \ +$$

$$P[G=T, S=F, R=T] \ +$$

$$P[G=F, S=F, R=F] \ +$$

$$= P[G=T \mid S=T, R=T] \cdot P[S=T \mid R=T] \cdot P[R=T] \ +$$

$$P[G=T \mid S=T, R=F] \cdot P[S=T \mid R=F] \cdot P[R=F] \ +$$

$$P[G=T \mid S=F, R=T] \cdot P[S=F \mid R=T] \cdot P[R=T] \ +$$

$$P[G=T \mid S=F, R=F] \cdot P[S=F \mid R=F] \cdot P[R=F] \qquad \textbf{Joint P.D.F}$$

$$= 0.01 \cdot 0.01 \cdot 0.2 + 0.01 \cdot 0.99 \cdot 0.2 + 0.01 \cdot 0.4 \cdot 0.8 + 0.4 \cdot 0.6 \cdot 0.8$$

$$= \textbf{0.1972}$$

$$\therefore P[R=T \mid G=T] = \frac{\sum_S P[G=T, S, R=T]}{\sum_{S,R} P[G=T, S, R]}$$

$$= \frac{0.002}{0.1972} \approx 0.01$$

Therefore the probability that *it is raining*, **given** that the grass is wet, is **1%**.

In the next section, we would be exploring different algorithms used to learn and construct belief networks. Learning belief networks plays a very important role as it enables us to perform *inferences* on the model and obtain valuable forecasts.

### 2.1.1.1   Learning Belief Networks

The sections demonstrate how belief networks represent a probability distribution over a set of variables, and how they can be used e.g. to predict variable states, or to generate new samples from the joint distribution. This section concerns on the process of obtaining a belief network, given a set of sample data. Learning a belief network can be split into two problems:

**Parameter learning:** Given a set of data samples and a DAG (Directed Acyclic Graph) that captures the dependencies between the variables, estimate the conditional probability distributions of the individual variables. Each factor is a conditional density depending on a restricted number of parameters which we can often estimate separately.

**Structure learning:** Given a set of data samples, estimate a DAG that captures the dependencies between the variables. This step determines which arcs are in the graph without looking at parameter estimates.

Most of academic literature focuses on structure learning which is the most challenging part. Indeed, the problem is certainly computationally expensive and a naive brute-force approach (greedy search) will usually not work. In what follows, we will describe the general framework of learning belief networks and present a general overview of two algorithms that can be used more specifically for structure learning.

### 2.1.1.1.1   Parameter learning

Let us assume that a graph $G$ has been selected and we have a parametric model $p(x_V; \theta)$, such that for each node $v \in V$, we can associate a subset of the parameters $\theta_{v|\mathrm{pa}(v)} \subset \theta$ and express the conditional density of $v$ given its parents,

$$p_{v|\mathrm{pa}(v)}(x_v; \theta) \;=\; p_{v|\mathrm{pa}(v)}(x_v; \theta_{v|\mathrm{pa}(v)}).$$

One of the major advatanges of using a graphical model is that the estimation problem is reduced to a set of lower dimension problems. The parameter learning step is computational feasible, assuming that $\theta_{v|\mathrm{pa}(v)}$, $v \in V$ form a partition of $\theta$. However, often it is enough to estimate the parameter $\theta_v$ in each component $p_{v|\mathrm{pa}(v)}(x_{v|\mathrm{pa}(v)} \mid \theta_{v|\mathrm{pa}(v)})$ separately, which could be done by standard estimators such as maximum likelihood estimator.

We can observe that the sparser the graph $G$ is, the smaller the expected dimension of parameters $\theta_{v|\text{pa}(v)}$, and therefore we can conclude that there is a computational incentive to select sparser graphs. We can also observe that issues arise when the sample size is smaller than the number of variables, which tends to increase the variability of data. This problem is frequently faced in Computational biology when the number of observations ends up smaller than the number of genes.

Parameter learning is of two main types:

**Maximum Likelihood Estimation:** A natural estimate for the CPDs is to simply use the relative frequencies, with which the variable states have occured. According to MLE, we should fill the CPDs such that $P(\text{data}|\text{model})$ is maximised. This is achieved when using the relative frequencies[25].

**Bayesian Estimation:** The Bayesian Parameter estimator begins with already existing prior conditional probability tables that express our beliefs about the variables before the data was observed.

**2.1.1.1.2 Structure learning** It is well known in the literature that the problem of learning the structure of belief is very challenging one to tackle: given computational complexity can often become super-exponential in the number of nodes in the worst case and polynomial in most real-world scenarios.

Structure learning is process of finding a directed acyclic graph (DAG) $G = \langle V, E \rangle$ that maximises $\mathbb{P}(G \mid D)$, and we can express the quantity as follows:

$$\mathbb{P}(\mathcal{G} \mid D) \;\; \propto \;\; \mathbb{P}(\mathcal{G})\,\mathbb{P}(D \mid \mathcal{G})$$

In context of Bayesian statistics, we could begin with selecting a prior $\mathbb{P}(\mathcal{G})$ over all possible DAGs[22]. In practice this is somewhat challenging since the set of all possible graphs grows super exponentially to the number of variables in the network. Without having any prior knowledge on the graph structure, it is commonplace to use the following uniform prior: for every pair $(i, j) \in V$, we assign:

- An edge $i \rightarrow j$ with probability $p_\ell = 1/4$, or

- An edge $j \rightarrow i$ with probability $p_r = 1/4$, or

- No arc between $i$ and $j$ with probability $p_0 = 1/2$

Another possibility is to assign all the three options with equal probability $p = 1/3$. Often in literature, a uniform prior is implicitly assumed with $P(\mathcal{G} \,|\, D) \propto P(D \,|\, \mathcal{G})$.

**2.1.1.1.2.1 Constraint-based learning** This class of algorithm [58] uses conditional independence tests to determine which arcs belong to the graph. Note that for that , we do not need to assume conditional independence in probability implies conditional independence on the graph so that $\perp\!\!\!\perp_p \Leftrightarrow \perp\!\!\!\perp_G$. Compared to the basic case, this is a stricter set of assumptions than the basic case where we assume $\perp\!\!\!\perp_G \Rightarrow \perp\!\!\!\perp_p$. The most popularly used independence tests are Pearson's $\chi^2$-test for categorical data, the exact $t$-test for Pearson's correlation coefficient (for Gaussian data), and Fisher's $Z$-test (for Gaussian data).

All constraint-based learning algorithms share a common three-phase structure as illustrated in the Inferred Causation algorithm. The first phase consists of learning the Markov blanket of each node to reduce the number of candidate DAGs early on. The second phase involves learning the skeleton of the DAG, that is, it identifies which undirected arcs are present in the DAG. Finally, in the third step, the arc directions are established and a complete partially directed acyclic graph is returned.

Before explaining the inferred causality algorithm, let us revise what Markov blankets are and their role in graphical models.

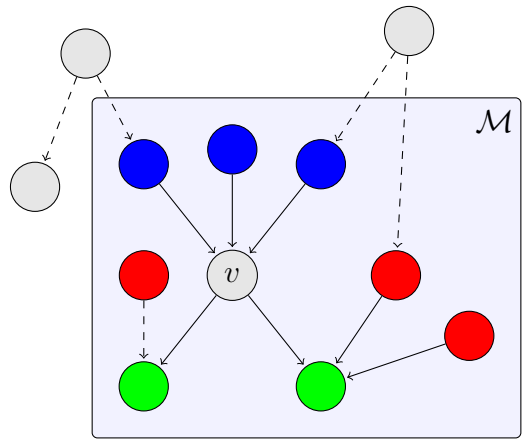

Figure 2.3: Illustration of the Markov blanket (shaded area) of a node in a simple DAG.

**Definition 2 (Markov blanket [53])** of a node $v$ is the union of the <span style="color:blue">parents</span> of $v$, its <span style="color:green">children</span> and its <span style="color:red">children's parents</span>. A useful property of the Markov Blanket is that every set of nodes is **conditionally independent** of $v$ when conditioned on the MB of $v$.

Unlike score-based structure learning that have countless research in optimisation theory and can be adapted to learning the network by using the the network score as the maximising objective function, constraint-based learning have few existing optimisation algorithms. The Inferred Causality algorithm leverages symmetries implied by the Markov blanket and provides a significant optimisation to the network learning process.

---

**Algorithm 1** Inferred Causality

---

**Input:** a data set containing the variables $X_i, i = 1, \cdots, m$.

**Output:** a completed, partially directed acyclic graph.

**Phase 1: learning Markov blankets**

1. For each variable $X_i$, learn its Markov blanket $\mathcal{B}(X_i)$.

2. Check weather the Markov blankets $\mathcal{B}(X_i)$ are symmetric i.e. $X_i \in \mathcal{B}(X_j) \leftrightarrow X_j \in \mathcal{B}(X_i)$, and drop of the asymmetric ones as false positives.

**Phase 2: learning neighbours**

3. For each pair $X_i, X_j, i \neq j$, search for a set $\mathbf{S}_{X_i X_j} \subset V$ such that $X_i \perp\!\!\!\perp X_j \mid \mathbf{S}_{X_i X_j}$ and $X_i, X_j \notin \mathbf{S}_{X_i X_j}$. If not found, place an undirected arc $X_i - X_j$. If $\mathcal{B}(X_i)$ and $\mathcal{B}(X_j)$ are available from from Phase 1, the search for $\mathbf{S}_{X_i X_j}$ can be limited to the smallest of $\mathcal{B}(X_i) \backslash X_j$ and $\mathcal{B}(X_j) \backslash X_i$.

4. Check weather the $\mathcal{N}(X_i)$ are symmetric, and correct asymmetries as in step 2.

**Phase 3: learning arc directions**

5. For each pair of non-adjacent variables $X_i$ and $X_j$ with a common neighbour $X_k$ such that $k \notin \mathbf{S}_{X_i X_j}$, set the directions of the arcs $X_i - X_k$ and $X_k - X_j$ to $X_i \rightarrow X_k$ and $X_k \rightarrow X_j$ to obtain a $\nu$-structure $\nu_l = \{X_i \rightarrow X_k \leftarrow X_j\}$.

6. Set the direction of the arcs that are still undirected by applying the following two rules recursively.

   a) If $X_i - X_j$ and there is a strictly directed path from $X_i$ to $X_j$, set the direction $X_i \rightarrow X_j$.

   b) If $X_i$ and $X_j$ are not adjacent and $X_i \rightarrow X_k$ and $X_k - X_j$, then set $X_k \rightarrow X_j$.

---

Although Inferred Causality is a popular method of constructing belief networks, score-based algorithms are equally popular too; each having a number of advantages over the other. In the next section, we will be exploring the process of learning a belief network using score-based learning and would be describing different optimisation techniques such as the Hill Climbing algorithm for constructing the belief networks.

**2.1.1.1.2.2  Score-based learning**  A popular method of constructing belief networks from data is the score-based approach[39]. The process assigns a score to each candidate belief network which quantifies how well a belief network $\mathcal{G}$ represents a dataset $\mathcal{D}$.

Assuming a structure $\mathcal{G}$, its score is given by,

$$Score(\mathcal{G}, \mathcal{D}) = Pr[\mathcal{G} \mid \mathcal{D}]$$

In other words, the posterior probability of $\mathcal{G}$ given the dataset. The score-based method attempts to maximise this score. Computation of the above formula can be succinctly cast into a more representative form by using Bayes Law:

$$Score(\mathcal{G}, \mathcal{D}) = Pr[\mathcal{G} \mid \mathcal{D}] = \frac{Pr[\mathcal{D} \mid \mathcal{G}] \cdot Pr[G]}{Pr[\mathcal{D}]}$$

Suppose that for each $v \in V$, an estimator $\hat{p}_{v|\mathrm{pa}(v)}$ of the corresponding conditional distribution is available. A common score to consider is the **Bayesian Information Criterion** (BIC).

The BIC seems to be biased towards the simpler structures, but as it gets more data it recognises that more complex structure is neccessary; in other words, it appears to trade off to data with model complexity, thereby reducing the extent of overfitting. The Bayesian Information Criterion is defined [21, 40] for a DAG $\mathcal{G}$ and data $\mathcal{D}$ as,

$$\mathrm{BIC}(\mathcal{G}, D) = \sum_{v \in V} \log \hat{p}_{v|\mathrm{pa}(v)} \left( X_{v|\mathrm{pa}(v)} \right) - \frac{1}{2} k_{v|\mathrm{pa}(v)},$$

where $k_{v|\mathrm{pa}(v)}$ is the parameter dimension $\theta_{v|\mathrm{pa}(v)}$.. For a discrete dataset, an estimator for a sample $\{X_V^{(i)}\}_{i=1}^n$ of size $n$ is given by,

$$\hat{p}_{v|\mathrm{pa}(v)} = \frac{\sum_i \mathbf{1}_{\{X_v^{(i)} = x_v, X_{\mathrm{pa}(v)}^{(i)} = \mathrm{pa}(v)\}}}{\sum_i \mathbf{1}_{\{X_{\mathrm{pa}(v)}^{(i)} = \mathrm{pa}(v)\}}}$$

Score-based algorithms attempt to optimise this score, returning the structure $\mathcal{G}$ that returns it. This can be a challenging task, since the space of all possible (undirected) structures is at least exponential in variables $n$, there are $\frac{n \cdot (n-1)}{2}$ possible undirected edges and $2^{\frac{n \cdot (n-1)}{2}}$ possible structures for every subset of those edges. Therefore, an Exhaustive search is unfeasible in time except in the most trivial of cases, $n > 6$. One possible choice in score-based structure learning is using Hill Climbing optimisation, as illustrated in Figure 2.4.

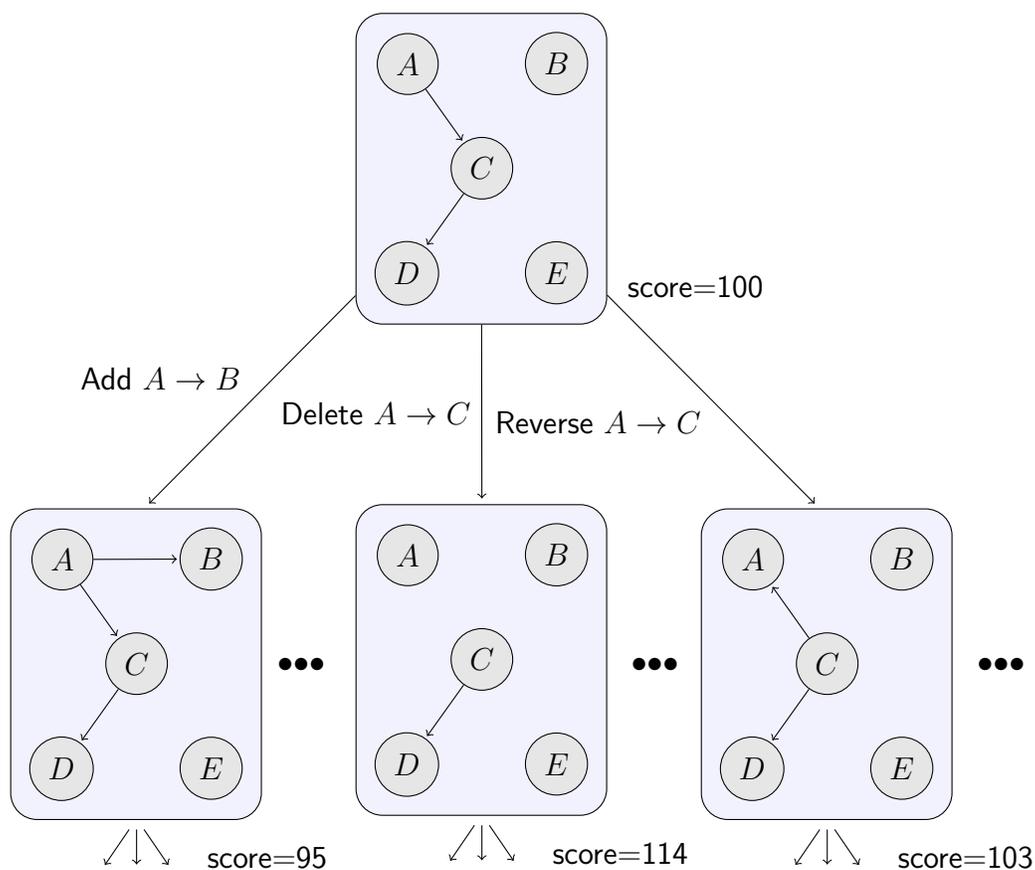

Figure 2.4: Illustration of a Belief Network structure hill climbing procedure[47].

Hill Climbing is an greedy, iterative, optimisation algorithm that is applicable to solving a vast array of problems. It starts with an initial, non-optimal solution to a problem, and then attempts to find the optimal solution to a problem by incrementally changing an element of the solution; if the change produces a better solution, an incremental change is made to the new solution. This process is repeated until no further improvements can be found.

Hill Climbing is also applicable in learning the structure (DAG) of a belief network [47]. We can define a **neighbourhood** in a DAG space as all networks we can reach by applying one of the three operators; adding an edge, removing and edge, and reversing the edge between two nodes[46]. The search is initiated from either an empty, full or possibly random network; although if there exists **expert** knowledge it can be used as a seed for the initial candidate network. At each step, we find the best neighbour and move to it. The algorithms main loop consists of attempting every single-edge operator, making the network that increases the score the most current candidate. The process halts when there is no single-edge change that increases the score.

There no guarantee that this algorithm will settle at a global maximum; it might settle at (different) local maximum if the climber starts at a poor location. A few solutions have been proposed to this problem, such as the avoiding structures in the `TABU` (a list of visited structures), and random restarts (applying random operators at random when at a local maximum).

There are a number of efficient implementations to the Hill Climbing algorithm for learning Bayesian networks. One such implementation proposes to use effectively cache relevant family scores in an AD-tree. The psuedocode of this implementation of this Hill Climbing algorithm[46] is given below.

---

**procedure** GREEDYHILLCLIMBING(`initial structure` $\mathcal{N}_{init}$, `dataset` $\mathcal{D}$, `scoring function` s, `stopping criteria` $\mathcal{C}$)

    $\mathcal{N}^* \leftarrow \mathcal{N}, \mathcal{N}' \leftarrow \mathcal{N}^*, \texttt{tabu} \leftarrow \{\mathcal{N}^*\}$

    **while** $\mathcal{C}$ is not satisfied **do**

        $\mathcal{N}'' \leftarrow \text{argmax}_{\mathcal{N} \in \textsf{neighbourhood}(\mathcal{N}') \text{ and } \mathcal{N} \notin \textsf{tabu}} \, s(\mathcal{N})$

        **if** $s(\mathcal{N}') > s(\mathcal{N}'')$ **then**           ▷ Check for local optimum

            $\mathcal{N}'' \leftarrow \texttt{random}(\mathcal{N}')$           ▷ Apply random operators

        **end if**

        **if** $s(\mathcal{N}'') > s(\mathcal{N}^*)$ **then**           ▷ Check for new best

            $\mathcal{N}^* \leftarrow \mathcal{N}''$

        **end if**

        $\texttt{tabu} \leftarrow \texttt{tabu} \cup \mathcal{N}'$

        $\mathcal{N}' \leftarrow \mathcal{N}''$           ▷ Move to neighbour

    **end while**

**end procedure**

---

Hill Climbing is not the only method for heuristic search. The best-first search (Russell and Norvig [63]), Dynamic Programming [46], and the max-min Hill Climbing Bayesian Network structure learning algorithm which combines both constraint-based and score-based methods [64]. However, score-based algorithms are considered more stable/robust compared to constraint-based algorithms and we would therefore be using them. We should note that for the Hill Climbing structure algorithm to be put in practice, we would have to build estimates for the probabilities, and would be using relevant estimators such as the **Maximum Likelihood Estimator** and **Bayesian Estimator**.

In the next section, we would be discussing the Markov chain, a graphical model which allows us to illustrate stochastic processes as a probabilistic finite-state machines.

### 2.1.2 Markov Chain

A MARKOV CHAIN is a graphical abstraction of a stochastic process that *transitions* between a finite number of *states*[27, 65]. A Markov Chain is primarily used to illustrate phenomenon where the future is conditional *only* on the present state and is *independent* of the past states [62]. This property, called *memorylessness*, is intrinsic to the definition of the Markovian Statistics and is formally referred to as the *Markov Property* in academic literature [48].

---

**Theorem 2 (Markov Property [57])** Given a *discrete-time* stochastic process $\{X_t, n \geq 0\}$, the future state $X_{t+1}, n \geq 0$ in a *countable set* $\mathcal{S}$ is conditional only on the present state $X_t$ and independent of the past states, $X_0, \cdots, X_{t-1}$[6]. More specifically, $\forall n \geq 0$, and $\forall i, j, k, \cdots, m \in \mathcal{S}$,

$$P[X_{t+1} = j \mid X_t = i, X_{t-1} = k, \cdots, X_0 = m] = P[X_{t+1} = j \mid X_t = i] = p_{ij}$$

---

Markov Chains can and are often illustrated as graphs (Figure 2.3 a). In the graphical model, each node represents a *state* and a *directed, weighted* arc for each non-zero transition probability (if $P_{ij} = 0$, the edge from node $i$ to $j$ is omitted). A number of disciplines in discrete mathematics, such as Graph Theory, are highly applicable in the graphical representation of Markov chains [57].

---

[6]This is assuming the discrete-time process is *first-order*. For $k^{th}$order Markov Chains, $1 \leq k \leq t$, the relationship is $P[X_{t+1} \mid X_t, X_{t-1}, \cdots, X_0] = P[X_{t+1} \mid X_t, X_{t-1}, \cdots, X_{t-k-1}]$

A *finite-state* Markov chain can also be represented as a matrix $[M]$ (Figure 2.3 b). Given the Markov chain has $\mathcal{N}$ states, $[M]$ is an $\mathcal{N} \times \mathcal{N}$ matrix $\{p_{ij}\}$. A number of disciplines within mathematics such as Linear Algebra, Probability Theory and Group Theory are applicable once a Markov chain is represented as a matrix [57].

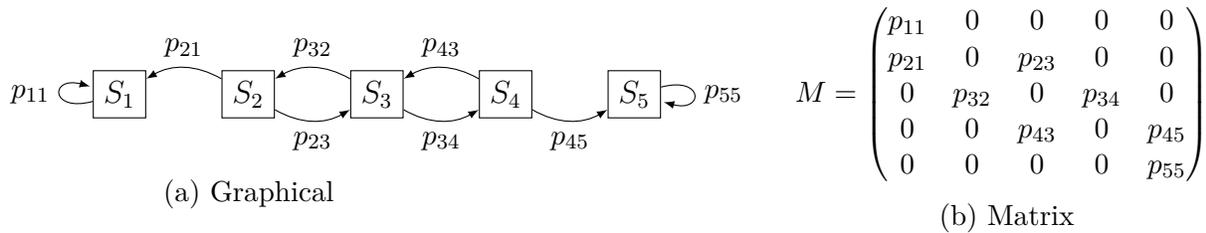

(a) Graphical

$$M = \begin{pmatrix} p_{11} & 0 & 0 & 0 & 0 \\ p_{21} & 0 & p_{23} & 0 & 0 \\ 0 & p_{32} & 0 & p_{34} & 0 \\ 0 & 0 & p_{43} & 0 & p_{45} \\ 0 & 0 & 0 & 0 & p_{55} \end{pmatrix}$$

(b) Matrix

Figure 2.5: Graphical and matrix representation of a 5-state Markov chain; a directed arc from $i$ to $j$ is included in the graph if and only if $p_{ij} > 0$ and $\forall i, \sum_j p_{ij} = 1$.

### 2.1.2.1 Hidden Markov Models

The HIDDEN MARKOV MODEL (HMM) is defined as a *generative* probabilistic graphic model, in which a sequence of *observable* symbols is generated by a sequence of *internal, hidden* states, which are not observed directly. The transition between the *hidden* states are assumed to be a *first-order* Markov chain as described in the previous section.

**Definition 3 (Hidden Markov Model [68])** is a 5-tuple $(Q, \sum, \Pi, A, B)$, where $Q = \{q_1, \cdots, q_N\}$ is a finite set of $\mathcal{N}$ states, $\sum = s_1, \cdots, s_N$ is the set of $\mathcal{M}$ possible symbols in the language, $\Pi = \{\pi_i\}$ is the initial probability vector, $A = \{a_{ij}\}$ is the state transition probability matrix, and $B = \{b_i(v_k)\}$ is the emission probability matrix. The HMM can be denoted by $\lambda = (\Pi, A, B)$, having the following constraints.

*The total transition probability [55, 68]*

- from the *initial* state to all *hidden* states is 1 i.e. $\sum_{i=1}^{N} \pi_i = 1$,

- from a *hidden state* $q_i$ to all other *hidden* states is 1 i.e. $\forall i \in Q, \sum_{j=1}^{N} a_{ij} = 1$,

- from the *hidden* states to all *observable* states is 1 i.e. $\forall i \in Q, \sum_{j=1}^{M} b_i(v_k) = 1$,

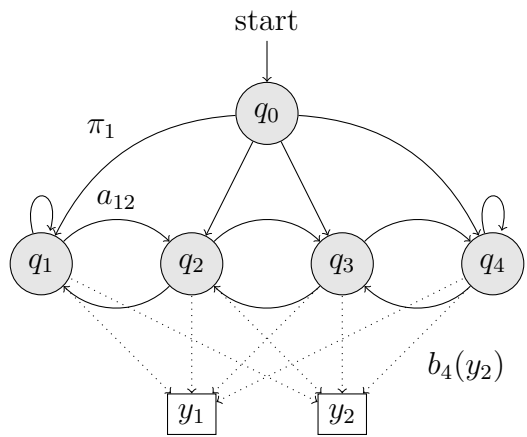

Figure 2.6: A 4-state HMM which can emit two discrete symbols $y_1$ or $y_2$. $a_{ij}$ is the probability to transition from state $s_i$ to state $s_j$. $b_j(y_k)$, is the probability to *emit* symbol $y_k$ in state $s_j$, and $\pi_i$ is the probability to transition from initial state $q_0$ to state $q_i$.

There are a number of problems associated with using an HMM [68]:

**Evaluation:** Evaluating the probability of an observed sequence of emissions $O = o_1 o_2 \cdots o_T$ ($o_i \in \sum$), given a particular HMM i.e. evaluating $p(O \mid \lambda)$.

**Decoding:** Determining the most likely state-transition path associated with an observed sequence $O = o_1 0_2 \cdots o_T$, i.e. evaluating $q^* = \mathrm{argmax}_q \ p(q, O \mid \lambda)$.

**Training:** Determining the ideal parameters for $\lambda$ to maximise the probability of generating an observed sequence of emissions $O = o_1 o_2 \cdots o_T$ ($o_i \in \sum$), i.e. evaluating $\lambda^* = \mathrm{argmax}_\lambda \ p(O \mid \lambda)$.

The solution to the first problem is given by the *forward and backward* iterative algorithms. The solution to the second problem is given by the famous *Viterbi* algorithm, which is a *dynamic programming* algorithm for finding the most likely sequence of hidden states given an observed sequence of emissions. Finally, the solution to the last problem is to use the *Baum-Welch* which calls both, forward and backward probabilties, to update the parameters iteratively.

**2.1.2.1.1 The Forward Algorithm** The FORWARD ALGORITHM [26] is used to calculate estimate the *optimal sequence of hidden states*, given both the *model parameters* and the *partially observed sequence*. In other words, it is used to calculate the *belief state*, the probability of a state at a given time, given the history of evidence (the partially observed sequence).

The forward variable defined as:

$$\alpha_t(i) = p(o_1, \cdots, t, q_t = s_i \mid \lambda)$$

i.e. the probability that the partially observed sequence up to time $t$ have been generated and the system is in state $s_i$ at time $t$. We can compute the $\alpha$'s using a recursive procedure:

**Base case:** $\alpha_1(i) = \pi_i b_i(o_1)$ (Having initial state as $q_0$ and generating $o_1$).

**General case:** $\alpha_{t+1}(i) = b_i(o_{t+1}) \sum_{j=1}^{N} \alpha_t(j) \cdot a_{ji}$, where $1 \leq t < T$ (Where $\alpha_1(i), \cdots, \alpha_T(i)$ corresponds to the $T$ observed symbols.

We observe that $p(O \mid \lambda) = \sum_{i=0}^{N} \alpha_T(i)$, given that any of $N$ states can be the halting state.

**2.1.2.1.2 The Backward Algorithm** The $\alpha$ values which were computed in previously using forward algorithms are applicable in solving the first problem, i.e. evaluating $p(O \mid \lambda)$. However, for training the HMMs, we require a second set of probabilities - the $\beta$ values [26].

Just like for the $\alpha$ values, we define the backward variable $\beta_t(i)$ as

$$\beta_t(i) = p(o_{t+1}, \cdots, o_T \mid q_t = s_i, \lambda)$$

i.e. the probability of the observed sequence from $(t + 1)$ to $T$ a given the state $s_i$ in time $t$. Just like the $\alpha$'s, we can compute the $\beta$'s using the following backwards recursive procedure:

**Base Case:** $\beta_T(i) = 1$ (Since we have no symbol to generate, we assume each state could be possible halting state).

**General Case:** $\beta_t = \sum\limits_{i=0}^{N} \beta_{t+1}(j) a_{ij} b_j(o_{t+1})$, where $1 \leq t < T$ ($o_{t+1}$ can be generated from any state $s_j$).

We observe that $p(O \mid \lambda) = \sum\limits_{i=1}^{N} \alpha_1(i)\beta_1(i)$, and as an inductive corollary, can be extended to $p(O \mid \lambda) = \sum\limits_{i=1}^{N} \alpha_t(i)\beta_t(i)$, where $1 \leq t \leq T$.

**2.1.2.1.3 The Viterbi Algorithm** The VITERBI ALGORITHM [26] is a dynamic programming algorithm that determines the most likely *state-transition* path given an observed sequence of symbols. Similar to the forward algorithm, the only difference is that we would be applying "max" rather than a "$\sum$", over all possible ways of arriving to the current state under consideration. It is an inductive algorithm in which every, at every instance you keep the observed sequence with the maximum probability for each of then $N$ states, as the intermediate state for the desired observation sequence $o = o_1, o_2, \cdots, o_T$.

Let $q = q_1 q_2 \cdots q_T$ be a sequence of states. We have to evaulate $q^* = \operatorname{argmax}_q p(q \mid O, \lambda)$, which essentially condenses to the problem of evaluating $q^* = \operatorname{argmax}_q p(q, O \mid \lambda) = p(q \mid O, \lambda) \cdot p(O \mid \lambda)$, and $p(O \mid \lambda)$, a constant, does not affect our choice of $q$ and therefore can be excluded.

The Viterbi Algorithm can be described as if "grows" the optimal path $q^*$ gradually while scanning every observed symbol. At time $t$, it keeps a record of *every* optimal path ending at each of the $N$ states.

Let $q_t^*$ be the optimal path for the subsequence of observed symbols $O(t) = o_1 \cdots o_t$ till time $t$, $q_t^*(i)$ being the path having the *maximum likelihood* ending at $s_i$, given the subsequence $O(t)$. Let $VP_t(i) = p\big(O(t), q_t^*(i) \mid \lambda\big)$ be the path of generating $O(t)$ by following the path $q_t^*(i)$. Therefore, $q_t^* = q_t^*(k)$, where $k = \arg\max_i \ VP_t(i)$ and $q^* = q_t^*$.

1. $VP_1 = \pi_i b_i(o_1)$ and $q_*^1 = (i)$.

2. $VP_{t+1}(i) = \max_{1 \leq j \leq N} VP_t(j) a_{ji} b_i o(t+1)$ and $q_{t+1}^* = q_t^*(k) \cdot (i)$, $1 \leq t < T$ where $k = \text{argmax}_{1 \leq j \leq N} VP_t(j) a_{ji} b_i \big(o(t+1)\big)$, where "$\cdot$" is a concatenation operator of states forming a path.

We can observe that $q^* = q_T^* = q_T^*(k)$, and $k = \text{argmax}_{1 \leq i \leq N} VP_T(i)$.

### 2.1.2.1.4 The Baum-Welch Algorithm

The previous problem of finding $\lambda^* = \text{argmax}_\lambda p(O \mid \lambda)$ is essentially a maximum likelihood estimation problem. If it were possible to observe the state transition path that had been followed when generating observed symbols, the estimation process would have simply condensed to counting the corresponding events and thereupon computing the relative frequency, however the state transition path is not observed. Therefore, in this situation, it is difficult to determine the maximum likelihood estimate analytically.

Fortunately, it is possible for us to use an *Expectation-Maximisation* (EM) algorithm for HMMs, called the BAUM-WELCH ALGORITHM [26]. Similar to other EM algorithms, we shall start with a random guess of the parameter values, and iteratively compute the expected probability of all possible transition paths and then re-estimate all parameters based on the expected counts of every corresponding event. This processes is repeated until the likelihood converges.

The process of updating formulae can be expressed in terms of the $\alpha$'s and $\beta$'s together with the current parameter values. We shall, however, introduce different notations two other notations. Let $\gamma_t(i) = p(q_t = s_i \mid O, \lambda)$ be the probability of being in state $s_i$ at time $t$, given $O$ (the observation sequence), and $\xi_t(i, j) = p(q_t = s_i, q_{t+1} = s_j \mid O, \lambda)$ as the transition probability from state $s_i$ to state $s_j$ given the $O$ observation sequence. We have

$\gamma_t(i) = \sum\limits_{j=1}^{N} \xi_t(i,j)$, and moreover, for $t = 1, \cdots, T$,

$$\gamma_t(i) = \frac{\alpha_t(i)\beta_t(i)}{\sum\limits_{j=0}^{N} \alpha_t(j)\beta_t(j)} \tag{2.1}$$

For $t = 1, \cdots, T-1$, $\xi_t(i,j)$ can be evaluated by:

$$\xi_t = \frac{\alpha_t(i)a_{ij}b_j(o_{t+1})\beta_{t+1}(j)}{\sum\limits_{j=1}^{N} \alpha_t(j)\beta_t(j)}$$

$$= \frac{\gamma_t(i)a_{ij}b_j(o_{t+1})\beta_{t+1}(j)}{\beta_t(i)} \tag{2.2}$$

The formulae for updating all parameters are:

- $\pi_i^{'} = \gamma_1(i)$

- $a_{ij}^{'} = \dfrac{\sum\limits_{t=1}^{T-1} \xi_t(i,j)}{\sum\limits_{j=1}^{N} \sum\limits_{t=1}^{T-1} \xi_t(i,j)}$

- $b_i^{'}(v_k) = \dfrac{\sum\limits_{t=1, o_t=v_k}^{T} \gamma_t(i)}{\sum\limits_{t=1}^{T} \gamma_t(i)}$

Graphical models are an interesting method of describing relationships between different factors affecting the particular system. Graphical models allow us to understand the relationships between different variables; for example we could use Kosaraju's algorithm [41] to find the strongly connected components and then use them for understanding the *connectedness* of a chain.

Having described the Graphical Models in depth, we would be using Python libraries such as pgmpy for implementing belief networks and we would be using hmms for implementing HMMs.

## 2.2   Python modules for Bayesian Programming

Python and R are one of the most popular programming languages which are used for probabilistic programming along with perhaps Julia and Stan. We would be solely using Python for the sake of simplicity, given that our model is macro-level and not a High Frequency Trading (HFT) and therefore we would only be back-testing on history, we would not need a low-latency, low-level programming language like C/C++ or perhaps even Java.

### 2.2.1   `pgmpy` : A Python Library for Probabilistic Graphical Models

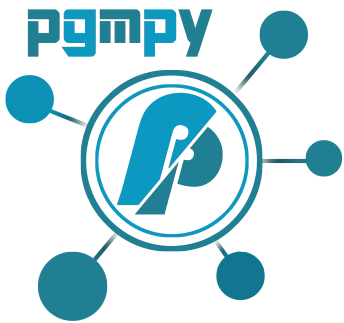

Figure 2.7: `pgmpy` is an open source Python library for graphical models.

`pgmpy` [3] is a Python library [2] for working with graphical models. It not only allows users to construct their own models but also allows them to answer *inference* or *MAP queries* over them. `pgmpy` has implementation of many inference algorithms such as Variable Elimination, Belief propagation, etc.

`pgmpy` was made by Ankur Ankan and Abinash Panda and was presented on Scipy Conference 2015 in Austin, Texas. We are highly grateful to both of them for writing an exhaustive documentation, supporting the project and being highly active on Gitter, answering queries of developers using `pgmpy` whenever they ran into a problem.

We would precisely be using `pgmpy` to construct the belief networks for the oil trading model. There are a number of methods for *learning* the belief networks by learning the data and we would be exploring such techniques offered by `pgmpy`. We would also be exploring how to use expert data from Energy Information Administration and the FRED and learn our model using that data.

### 2.2.2 `hmms`: A Python Library for Hidden Markov Models

`hmms` is an open-source Python library for Hidden Markov Models (HMMs). It is a general purpose, easy to use library which implements important methods needed for the training, examining, and experimenting with Hidden Markov Models. The library was created by Luká Lopatovský as part of his thesis at Czech Technical University, Prague [45]. The library was created with an objective of having a wide user base, and for that reason, it was chosen to be in Python. However, some of its computationally expensive modules have been written in Cython and some also use function calls from SciPy and NumPy.

`hmms` provides interactive code examples as `iPython` notebooks, with library usage examples that cover all the main uses of the library. A user can easily follow the notebook and get an understanding of how to use the library and implement Minimum Working Examples of the features provided by `hmms`.

`hmms` offers implementations for constructing Discrete Time Hidden Markov Models (DtH-MMs), with the initialisation being simply passing three parameters $\theta = (Q, A, \Pi)$ as `numpy` arrays. It also allows generating random state and emission sequences (using parameters of a model), calculating probability of state and emission sequences, and saving and reading parameters from a file. Moreover, it also provides implementations for the Viterbi algorithm and Baum-Welch algorithm, which we would be using for time-series analysis. `hmms` also provides functionality to dealing with a Continuous Time Hidden Markov Model (CtHMM), and run multiple trains by one function call.

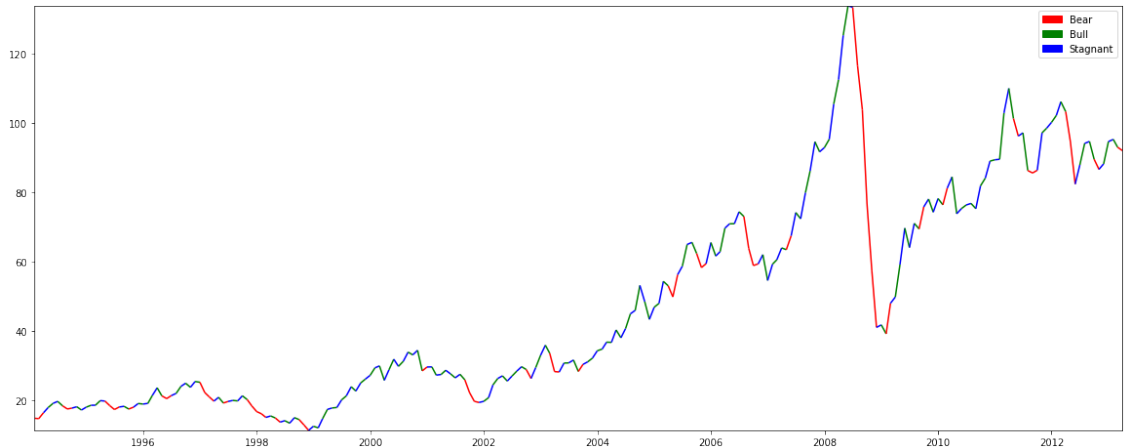

Figure 2.8: `hmms` provides a graphical representation of HMMs, illustrating the hidden states and emission states as state and emission sequences.

Given that Hidden Markov Models are highly applicable in a number of time-series analysis applications (Barber, D) [6], we would be using `hmms` for analysing financial time-series data obtained from the FRED and EIA. We would be using `hmms` to identify bull and bear regimes in time-series data in an attempt to discretise our time-series to be used in a Bayesian Network. We would be discussing this in the next chapter.

## 2.3   Datasets in macroeconomic perspective

We would be using the Energy Information Administration (EIA) and the Federal Reserve Economic Data (FRED) to retrieve datasets of different macro-economic data to incorporate in our model. Fortunately, both data centres have Python APIs and allow to neatly retrieve data as a `pandas` dataframes without any messy `curl` or `wget` UNIX calls.

### 2.3.1   Physical market factors from EIA

The Energy Information Administration (EIA) has a number of datasets for macro-economic inference on the oil markets. Other than by collecting macro-economic data which affects the oil markets, the EIA also assists oil market analysts in constructing statistical models for the oil markets, along with providing an expert knowledge about the relationship between the different macroeconomic variables.

In a report compiled by EIA [7] Kilian, Lutz, short-term oil prices have been forecasted using vector autoregression (VAR), spread between oil future prices and oil spot prices, forecasts based on non-oil industrial commodity prices, and the forecasts relying on the time-varying parameter (TVP) model of the spread between U.S. gasoline spot prices and heating oil, and spot price of crude oil.

In a study [9] conducted by Bruce Bawks, a financial analyst of the Energy Markets at the EIA, it was determined that several macroeconomic factors affect the price of oil and therefore these factors could be used in models for a short-term forecast of oil markets. The price of oil can be influenced by geopolitical events, such as OPEC and non-OPEC relations, and events resulting in political instability in the Middle East such as the Iran-Iraq War and Iraq's invasion of Kuwait. Market inefficiency [7]in oil markets can allow arbitrage, therefore affecting the price in different markets. Economic growth has a strong affect on the price of oil (which would be discussed in FRED data). Changes in OPEC production, such as unplanned supply disruptions or production quotas, affect the price of oil. Determining every factor which affects the price of oil is unfeasible, given F.A Hayek's notion of that no single agent in the market can determine all factors affecting the price and production in a system [34].

The relevant macroeconomic data from the EIA is updated at a monthly frequency. This results in a constraint on the time period at which we get predictions on statistical confidence. Delaney Mackenzie at Quantopian, described what is known as the "30x rule of thumb" , which asserts that if the data is sampled at frequency $f$, we can only achieve statistical confidence at $30f$. So given we have data sampled at a monthly frequency, we only can achieve predictions on a multi-year trend.

---

[1]Inefficient markets refer to the phenomenon where an asset's market price does not accurately reflect its true value, therefore allowing traders to simultaneously purchase and sell the asset for a profit.

## 2.3.2   FRED

The Federal Reserve Economic Data, St. Louis (FRED) is a collection of 508,000 US and international time series from 86 sources [60] on different macroeconomic data such as the consumer price index, producer price index, and Industrial Production index. Similar to the EIA, the FRED not only provides macroeconomic datasets but also provides expert knowledge about the relationships between different sectors of the economy. The FRED also provides a lot of expert knowledge about how different macro-economic indicators play a role in determining the price of oil.

In a study by Kilan, Lutz study [38], a structural decomposition of the real price of crude oil was proposed; the three factors affecting the price of oil were crude oil supply shocks, shocks to the global demand for all industrial commodities, and demand shocks that are specific to the crude oil market. Basing on Kilians study, Alejandro Badel and Joseph McGillicuddy [4] at the FRED carried out a research, exploring the correlation between the oil prices and inflation expectations, and came up with three conclusions.

Firstly, the study concluded the fact that in the entire time frame 2003-2015 there was little correlation between the inflation expectation and oil price, but during the 2007-2008 Financial Crisis there was a high correlation suggests that the financial crisis was some sort of an 'exception' and the level shift in the inflation expectations unrelated to the oil price shocks. Secondly, it concluded that only the correlation of oil-specific demand shocks and inflation expectations was positively significant across all sub-periods (excluding the 2007-2008 Financial Crisis) Thirdly, it suggests that there is a need to investigate the nature of oil specific demand shocks.

Similar to the EIA data, the macroeconomic data from the FRED comes at a monthly frequency, rendering our model accurate to multi-year trends (30 months) by the *30x rule of thumb*. In a research conducted by Kilian [8], it has been concluded that not has been lost by not using high-frequency data, however, given that MIDAS model was used, it might be possible that HMMs are better trained with more high-frequency data.

The FRED, just like the EIA fortunately has a (third party) Python API [35] which allows us to retrieve data very, very neatly as a `pandas` dataframe from the FRED database.

Similar to the EIA data, FRED data often has holes, however, the indexes are formatted as `dates` so thus not much data preprocessing would be needed.

# Chapter 3

# Design

In this section, we would be describing the design process of our oil trading model. The proposed design process is based on an earlier research used to quantify public bicycle behaviour in Xian [66]. Laying out a clear design plan will allow us to not only describe the structure of our model, but would also allow us to have clear outline of the sprints required hence lay a solid foundation to the implementation stage.

## 3.1   Macro Strategy

Most speculation on oil is done on the interpretation and prediction of large-scale events relating to geopolitics, monetary policy, privatizations of national petroleum and gas companies, supply and demand factors, and natural disasters - thus making oil trading a global macro strategy [29]. Macroeconomic indicators are used as buy and sell signals on a quarterly (or perhaps even monthly) scale.

These macroeconomic variables are usually in the form economic time-series data provided by federal statistical bodies responsible for the collection, analysis, and dissemination of macroeconomic data. We would be using the Energy Information Administration (EIA) for data for the energy markets and we would be using the Federal Reserve Economic Data (FRED) for macroeconomic data.

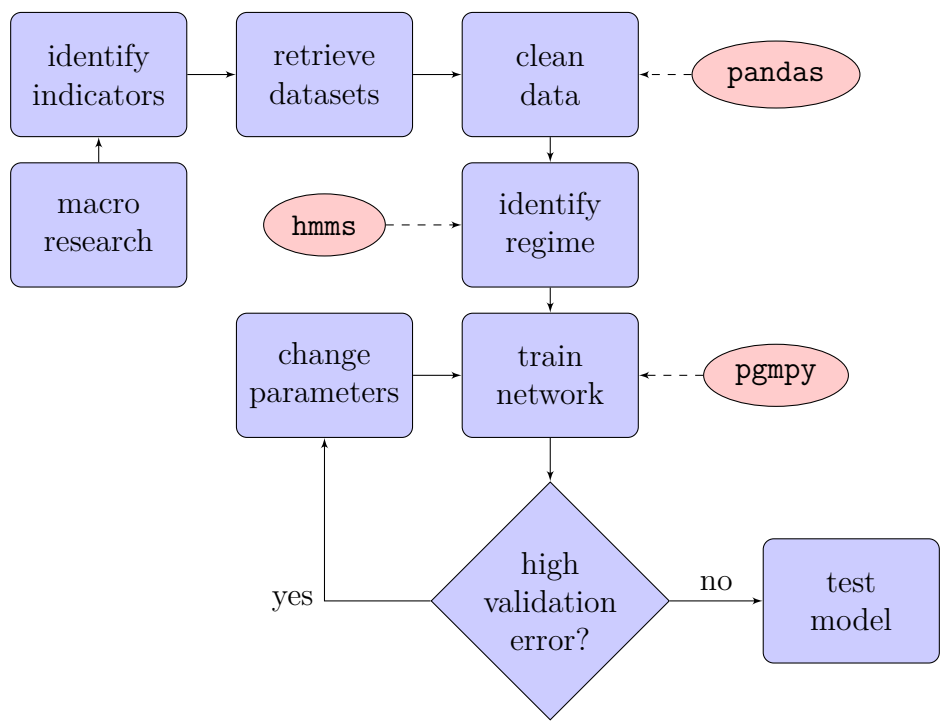

Figure 3.1: An illustration of the design process for the Crude Oil Trading model.

## 3.1.1   Energy Information Administration

Other than providing time-series data on the crude oil markets markets, the EIA also provides expert economic knowledge of the structure of the oil markets, latest news and reports of the energy markets, and forecasts of these macroeconomic variables. We would be priciesly using these forecasts as inputs to for inference on our learnt model to obtain the forecast of the price of oil as the output.

The EIA provides a number of datasets which are easily accessible from their open-data facilities. These datasets, such as the Annual Energy Outlook, International Energy Data, and the Short-Term Energy Outlook, provide vital information on the physical market factors of the oil markets. The Short-Term Energy Outlook (STEO) contains datasets which have forecasts for the supply and demand of crude oil amongst forecasts for many other factors affecting the energy sector. We would be mostly be using data from STEO to learn our model and would be using the forecasts as inferences to our model to make predictions.

The EIA assess the various factors that influence the crude oil prices; physical market fac-tors as well as those relating to trading and financial markets [12]. The EIA lists non-OPEC and OPEC crude oil production, non-OECD and OECD consumption of crude oil, OECD inventories, proven reserves, geopolitical events, and the trading of oil in financial markets as the main factors affecting the price of oil. We would be disseminating these factors and understanding what role they play in the crude oil markets.

### 3.1.1.1  Supply Non-OPEC

Crude Oil production from countries outside the Organization of the Petroleum Exporting Countries (OPEC) currently represents about 60 percent of world oil production. The way the oil producers operate in non-OPEC countries is fundamentally different to how they operate in the OPEC countries in a number of ways [13]. In contrast to OPEC oil producers which have a centralised decision making body, non-OPEC producers make independent decisions about oil production. Whereas OPEC producers are mostly in the hands of nationalised oil companies (NOCs), non-OPEC producers are usually international or investor-owned oil companies (IOCs). Unlike the OPEC producers, non-OPEC producers are price takers; they respond to market prices than trying to influence the oil prices by managing production [18]. Ceteris paribus, when there is a non-OPEC supply cut, there is an upward push on prices and an expectation from OPEC to fill the supply void.

### 3.1.1.2  Supply OPEC

Crude Oil production [14] in the Organization of the Petroleum Exporting Countries (OPEC) [14] plays an pivotal role [59] in the stability of the oil markets. Given that it holds 40 percent of the crude oil production and 60 percent of the petroleum traded internationally, the cartel actively seeks to control oil production in its member countries by setting production targets. Changes of crude oil production from the OPECs largest producer, Saudi Arabia, often affects the price of oil. OPEC producers maintain spare capacity of producing oil which is often used as a tool for influencing the markets. Despite OPECs efforts to manage production to maintain targetted price levels, OPEC member countries do not always comply with production targets adopted by the cartel, as they are often not in the interests of some member countries.

### 3.1.1.3 Demand OECD

The Organization of Economic Cooperation and Development (OECD) consists of the United States, much of Europe, and other developed countries such as Turkey. These developed economies consume more oil than the non-OECD countries, but have much lower oil consumption growth. Oil consumption in the OECD countries actually declined in the decade between 2000 and 2010, whereas non-OECD consumption jumped to 40 percent during the same period.

Structural conditions in OECD member countries economies influence[11] the relationship between oil prices, economic growth, and oil consumption. For example, economies in OECD countries tend to have larger technology and services sectors relative to industrial manufacturing. Therefore economic growth in these countries may not have the same impact on oil consumption as it would in non-OECD countries. Moreover, given the available public transportation in OECD countries, consumers have an easy alternative to privately owned vehicles in the face of increasing oil prices.

### 3.1.1.4 Demand Non-OECD

Oil consumption in developing countries outside the Organization of Economic Cooperation and Development (OECD) has risen sharply in recent years.While the OECD member states' oil consumption was declining from 2000 to 2010, non-OECD consumption increased by almost 40 percent [10].

Structural conditions in non-OECD countries are entirely different to that in OECD countries and influence the relationship between oil prices, economic growth, and oil consumption in an entirely different manner. Developing countries tend to have a greater proportion of their economies in manufacturing industries which are relatively more energy intensive than service industries. Due to lack of public transport infrastructure, vehicle ownership per capita is highly correlated with rising incomes and has much room to grow in non-OECD countries therefore pushing the oil price upwards by increasing oil consumption.

### 3.1.1.5 Balance

Inventories [67] play an important role in the oil markets because they act as the balancing point between supply and demand. During periods of over-production, crude oil and petroleum products can be stored for future use and similarly, during periods of over-consumption, crude oil and petroleum products can be used from the inventory. During the 2008 Financial Crisis, the unexpected drop in crude oil consumption led to record crude oil inventories in the United States and other OECD countries. Given the uncertainty in the supply and demand, petroleum inventories are often seen as a precautionary measure.

For quantifying balance, we would be using `STEO.PASC_OECD_T3.M` which represents the monthly OECD End-of-period Commercial Crude Oil and Other Liquids Inventory.

### 3.1.1.6 Other EIA data

Other than using the seven factors [20] idenfied by the EIA that drive the price of crude oil, we would be using other macroeconomic data chosen by trial and error that increases the accuracy of our model. We would be using datasets such as the U.S. Crude Oil market physical factors [32] and U.S. foreign exchange rate [16].

| Series ID | Frequency |
|---|---|
| STEO.RGDPQ_NONOECD.M | M |
| STEO.RGDPQ_OECD.M | M |
| STEO.PAPR_NONOPEC.M | M |
| STEO.PAPR_OPEC.M | M |
| STEO.PATC_OECD.M | M |
| STEO.PATC_NON_OECD.M | M |
| STEO.COPRPUS.M | M |
| STEO.CORIPUS.M | M |
| STEO.FOREX_WORLD.M | M |
| STEO.PASC_OECD_T3.M | M |
| STEO.COPS_OPEC.M | M |
| STEO.COPC_OPEC.M | M |
| STEO.T3_STCHANGE_OOECD.M | M |
| STEO.T3_STCHANGE_NOECD.M | M |

Table 3.1: Datasets obtained from the EIA.

### 3.1.2 Federal Economic Reserve Data

As a matter of fact, we are also using data from the FRED, which is one of the most primary data sources for macroeconomic data. There are a number of macroeconomic variables which play a pivotal role in the oil markets and the price of crude oil has long been influenced by the monetary policy dictated by the Federal Reserve Bank [33], the macroeconomic atmosphere [28] of the economy, industrial growth, industrial production, capacity, and capacity utilization.

| Series ID | Frequency |
|-----------|-----------|
| CPIENGSL | M |
| CAPG211S | M |
| CAPUTLG211S | M |
| IPG211S | M |
| IPG211111CN | M |
| INDPRO | M |
| IPN213111N | M |
| PCU211211 | M |

Table 3.2: Datasets obtained from the FRED.

## 3.2 Preprocessing data (Cleaning data)

Data preprocessing is the process by which data inconsistencies such as missing values, noise, out-of-range values, and unformatted data are removed so that the data can be used with `hmms` and `pgmpy`. Analysing data that has not been carefully preprocessed can often lead to harmfully misleading results. Data preprocessing is therefore one of the most elementary steps towards harnessing the power of a dataset, however it is also one of the most time-consuming processes which involves alot of experimentation with trying diffrerent data structures and different data analysis tools. The final product of data preprocessing is the **training** dataset, which would be used to train our model to be used on the **testing** dataset for assessing the performance of our model.

As described earlier, there are a number of available Python libraries for retrieving data from the EIA and FRED and for data preprocessing. Given the data retrieved from the EIA[36] and FRED[35] are often unable to be used directly due to their unstructured for-mat, we would be using `pandas`[52] for data preprocessing and noise-reduction, respectively, so that they could be transformed into formats which could be used with both, `pgmpy` and `hmms`. In the next subsections, we would be describing the tools and the methods being used for data preprocessing in this project.

## 3.2.1   `pandas` - powerful Python data analysis toolkit

`pandas` is an open-source Python library that provides fast, flexible, and expressive data structures designed to make working with "relational" or "labeled" data both easy and intuitive. It aims to be the fundamental high-level building block for doing practical, real world data analysis in Python. In many ways, `pandas` can be compared to SQL in that it provides functionality to construct dataframes (synonymous to tables) and provides functionality to carry out queries on the data.

Data obtained from the EIA and FRED often at times are replete with missing values (`NaN`, `inf` values), have inconsistent date ranges for the required date ranges, have indices which are formatted as strings rather than the Python `datetime`, and have intersecting date ranges and congruent frequencies, yet different dates (so indices of all `pandas.Series` in a dataframe have to be made the same for them to be merged).

Quantopian, a crowd-sourced hedge-fund for freelance quantitative analysts, provides academic resources for prospective quants. Amongst these lectures was an exhaustive lecture by Maxwell Margenot [49] on `pandas`, which was highly used in the data preprocessing. We would be using `pandas` to solve all the data inconsistencies and would consequently obtain the **training, validation and testing** dataset so that it could be used with `pgmpy`.

## 3.2.2   Obtaining the training, validation, testing set

In Machine Learning projects, it is almost always required to construct algorithms that learn and make predictions based on data. In our project, we would be making data driven decisions by inputing our data in a Bayesian model.

In order to learn the belief network, we would be using the $k$-**fold** cross validation method, with $k = 3$. In the $k$-fold cross-validation method, the data are is partitioned into $k$ subsets. Each subset is used in turn to validate the model fitted on the remaining subsets. In our example, we would be sticking to the **leave-one-out cross-validation technique**, thus having three datasets; the training set, the validation set, and the testing set.

**Training set:** is a sample used for learning the Belief Network using Hill Climbing.

**Validation set:** is a sample used to tune the parameters of the belief network. In our example, we experimented with using different *Bayesian Dirichlet* scoring functions such as the `BDeu` or `K2` and the Bayesian Information Criterion (`BIC`)

**Testing set:** is a sample used only to assess the performance of a **fully-trained** model. In our project, we would be using the testing dataset to estimate the error rate after we have chosen our final model with the scoring function that gives the most accurate validation dataset.

The difference between the validation and testing datasets is often at times vastly misunderstood and both datasets are often incorrectly equated. The validation dataset is used for model selection and the test set is used for assessing the error in the final model that was selected to be "tuned" with the validation set[30]. A popular ratio of $80 : 10 : 10$ is used for spliting the data and we would be adhering to this standard. Moreover, given that the validation dataset and the testing data set would each be of 33 months (recall the "30x rule of thumb") hence being a appropriate length of time to have statistical confidence at.

However, before we decide to split our preprocessed data into three datasets, there is a vital step that needs to be carried out. The `pandas.DataFrame` that `pgmpy` takes in as input has values of a **discretised** nature i.e. having a **fixed** number of states. The data obtained from the EIA and FRED is raw continious time-series data which has almost an infinite number of states. In order for `pgmpy` to use this data, it is essential we find a mechanism to find **hidden** states in time-series data for discretisation of this data. We would be describing the process of discretising the EIA and FRED data in the coming section.

## 3.3 Time-series discretisation

Belief Networks come handy when representing single independence models with *discrete* states, however they do not allow us to model changes of random variables over time. Due to such changes in *regimes*, it is necessary we use different belief networks at different times in order to have an appropriate model over random variables . In this section, we propose using Hidden Markov Models to *discretise* the time-series data so that they can be used in a belief network.

The general idea of **systematic** oil trading is to use **macroeconomic** and **physical market** data to identify the **structure** of the markets, thus assisting us in identifying opportune times to be *long* in the oil markets, and periods when it is beneficial to be *short*. As a matter of fact, it is obvious that we would prefer to be *long* before periods where oil markets are in a **bull regime** and be *short* before periods of a **bear** regime.

### 3.3.1 Regime Detection

Regime detection is the process by which we identify periods of similar volatility in a time-series model. These periods could either be bull/bear regimes, periods of low or high volatility, or perhaps some other characteristic that cannot be qualitatively described. Given that these periods are **latent** in nature and can only be **observed** as **emissions** (returns), the problem ultimately decomposes to the problem **Hidden Markov Models** were constructed to solve.

In our design specification, we have decided to use Hidden Markov Models[1]to identify these regimes in the time-series data. The HMM would be a 5-tuple $(Q, \sum, \Pi, A, B)$, and the regime detection will comprise of four main stages,

**Transforming** the time-series into an **emission sequence** by *first-order intergration* of the concerned time-series and replacing the *returns* with the **parity** of the returns.

**Learning** the parameters of the assumed HMM generating the time series by using the **Baum-Welch algorithm**, hence allowing us to obtain the $\Pi$ (initial state probabilities), $A$ (state transition probability matrix), and $B$ (emission matrix).

**Finding** the most likely sequence of *hidden* states as the underlying regimes that possibly generated the sequence of *observed* emissions using the **Viterbi algorithm**.

**Identifying** the *latent* meaning behind each *hidden* state by indexing them according to their *arthimetic mean.*

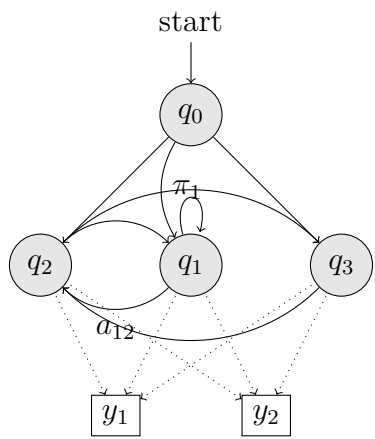

Figure 3.2: A graphical illustration of the proposed 3-state HMM generating the behaviour of a financial time-series.

The Python implementation for HMMs provided by `hmms`[43] is perhaps one of the most easy to use, given its concise yet exhaustive documentation as an iPython notebook[44], and its *ready-to-use* implementations of the **Forward-Backward**, **Viterbi** and **Baum-Welch** algorithms, amongst many other functionalities.

## 3.4 Constructing structure of the Crude Oil Markets

Belief networks are highly applicable to capture **causal** relationships between both; standard *data-driven economic variables*, and *quantified expert judgements* about the geo-politics of the oil market (in particular the production and capacity policy of OPEC members). Given the exponentially increasing rate of data collection, it is almost infeasible for oil speculators to construct belief networks alone by expert knowledge. Moreover, given the dynamic nature of the structure of the commodities markets, minor structural changes are always taking

---

[1]Applying HMMs to regime detection often can be a challenging, as there is in particular no known method by which we can determine the *meaning* of a state, or even the number of states.

place and therefore it is infeasible to keep update with the structure of the markets.

We therefore propose using structure learning algorithms for constructing a *causal* network of the oil markets. In Chapter 2, we had discussed two main structure learning algorithms; the **constraint-based** Structural learning, and **score-based** Structural learning. Despite recent advances in constraint-based learning algorithms[54], we would be using score-based algorithms, which have benefited a lot from past and recent research in optimization theory, using the network score as the objective function to maximise[58]. A popular score-based structure learning algorithm is **Hill Climbing**, a greedy, iterative search algorithm used to maximise the network score, which we would be using to learn the structure of the oil markets and perform inferences to make forecasts of the future.

However, we would not be using Hill Climbing to learn the entire structure of the oil markets. Using Hill Climbing without any initial structure not only results in models which might be *causaly* incorrect, but could also result in an rather drastically different network whenever constructed again during the *validation* step. This is due to the nature of the Hill Climbing algorithm, which is highly affected from where the initial point is taken and can get "stuck" at a number of different *local* maxima rather than the one *global* maxima. As a result, we have decided to use an expert drafted network having a basic structure of the oil markets, constructed by the expert using knowledge of laws of supply and demand in economics. Though there is chance that the re-learnt network during the validation step might be different, chances are less that the changes will not be entirely random and the basic expert framework would still exist.

We would be using `pgmpy`, which provides a very straightforward implementation to the **Hill Climbing** algorithm. `pgmpy` provides an excellent notebook [1] for learning belief networks. We would be using our training data to **estimate the structure of the network**, and would then fit the data using an **estimator** to learn the parameters. Below is a minimal working example in Python. `pgmpy` also provides an excellent facility of using an initial structure to learn the network from, and we would be using the network constructed by the expert as our initial model.

Below is a Minimal Working Example of implementing a Hill Climb search on (randomly) produced data, using `pgmpy`.

---

**Listing 1** Minimal working example

```python
import numpy as np
import pandas as pd
from pgmpy.estimators import BicScore
from pgmpy.estimators import HillClimbSearch
from pgmpy.estimators import BayesianEstimator

# Generate (discretised) data with dependencies

data = pd.DataFrame(np.random.randint(0, 3, size=(2500, 8)),
                    columns=list('ABCDEFGH'));

data['A'] += data['B'] + data['C'];
data['H'] = data['G'] - data['A'];

data_train = data[: int(data.shape[0] * 0.75)];

# Learn network structure

hc = HillClimbSearch(data_train, scoring_method=BicScore(data_train));
model = hc.estimate();

# Learn parameters of the network

model.fit(data_train,
          estimator=BayesianEstimator, prior_type="BDeu");

# Test the dataset

data_test = data[int(0.75 * data.shape[0]) : data.shape[0]];
data_test.drop('A', axis=1, inplace=True); # Drop variable to be predicted
prediction = model.predict(data_test); # Obtain prediction
```

---

## 3.5 Assessing the model

In Machine Learning projects, it is almost always required to construct algorithms that learn and make predictions based on data. In our project, we would be making data driven decisions by inputing our data in a Bayesian model.

In order to learn the Bayesian Network, we would be using the **Leave-$k$-out** cross validation method. In the $k$-fold cross-validation method, the data are randomly partitioned into $k$ subsets. Each subset is used in turn to validate the model fitted on the remaining $k - 1$ subsets. In our example, we would be sticking to the leave-one-out cross-validation technique, thus having three datasets; the training set, the validation set, and the testing set.

**Training set:** is a sample used for learning the Belief Network using Hill Climbing.

**Validation set:** is a sample used to tune the parameters of the belief network. In our example, we experimented with using different *Bayesian Dirichlet* scoring functions such as the `BDeu` or `K2` and the Bayesian Information Criterion (`BIC`)

**Testing set:** is a sample used only to assess the performance of a **fully-trained** model. In our project, we would be using the testing dataset to estimate the error rate after we have chosen our final model with the scoring function that gives the most accurate validation dataset.

The difference between the validation and testing datasets is often at times vastly misunderstood and both datasets are often incorrectly equated. The validation dataset is used for model selection and the test set is used for assessing the error in the final model that was selected to be "tuned" with the validation set[30]. A popular ratio of `80 : 10 : 10` is used for spliting the data and we would be adhering to this standard. Moreover, given that the validation dataset and the testing data set would each be of 33 months (recall the "30x rule of thumb") hence being a appropriate length of time to have statistical confidence at.

However, before we decide to split our preprocessed data into three datasets, there is a vital step that needs to be carried out. The `pandas.DataFrame` that `pgmpy` takes in as input has values of a **discretised** nature i.e. having a **fixed** number of states. The data obtained from the EIA and FRED is raw continious time-series data which has almost an infinite number of states theoretically having a range of $\mathbb{R}^+$. In order for `pgmpy` to use this data, it is essential we find a mechanism to find **hidden** states in time-series data for discretisation of this data. We would be describing the process of discretising the EIA and FRED data in the coming regime.

In the next chapter, we would be implementing the design process as described in Python in an attempt to experiment with the proposed models and observe their performance, feasability, reliability, and return.

# Chapter 4

# Implementation

**(Using Probabilistic Graphical Models to forecast the price of crude oil)** The document contains the **implementation** of the oil trading system using graphical models.

- Section 1 is dedicated to retrieving data from the **EIA** and **FRED**, **preprocessing** the data, and creating the **training**, **validation**, and **test** datasets.
- Section 2 is dedicated to implementing a **regime detection model** using **Hidden Markov Models** to identify bull, bear, and stagnant regimes.
- Section 3 is dedicated to learning the **macroeconomic structure** of the oil markets by **learning the belief network** using **hill-climbing** structural learning.
- Section 4 is dedicated to **testing** the constructed model by simulating trades and taking positions based on those trades.

It is recommended to take look at the thesis documentation to understand the basis on which we selected the macroeconomic economic data from the EIA and FRED, and the theoretical context of the graphical models being employed in our model.

We would be using a number of Python packages, such as pgmpy and hmms throughout the notebook and it is highly recommended that we take a deeper look at them as only a specific and relevant functionality of those packages have been used in our model.

## 4.1 Data preprocessing

Data preprocessing plays an important role in Machine Learning. Our data preprocessing has four main steps; data retrieval, data cleaning, data transformation, and data discretisation.

### 4.1.1 Data retrieval

The first step of constructing our model is to retrieve the data from the open-data facilities. We have selected the **EIA** and **FRED** as our primary data sources. Unfortunately, both these open-data facilities do not provide Python packages to neatly retrieve data in Python, so we will have to resort to using third-party APIs. For the EIA, we are using EIA-python, and for the FRED we are using fredapi.

Before beginning, we would be first be importing pandas and numpy, as they are highly required in the entire data preprocessing section.

```
In [1]: import pandas as pd
        import numpy as np
```

We would now be retrieving data, beginning with the EIA. Before retrieving data from the EIA, we have to register with EIA's open-data facility, in return of which we shall recieve an API key, which is used as a passphrase to access data from the EIA's datacenter.

```
In [2]: # Importing the library
        import eia
        # the API key we recieved from EIA
        eia_key = "XX";
        # Initiates a session with the EIA datacenter to recieve datasets
        eia_api = eia.API(eia_key);
```

Now, we shall be making a request to retrieve data from the EIA as a pandas dataframe. EIA provides a 3,872 Short-Term Energy Outlook (STEO) datasets, with short-term (2-year) forecasts of each dataset. These datasets can be searched in EIAs query browse facility, which also offers a catalogue of different datasets sorted by relevance. Just as an example to demonstrate, we would be retrieving the **Crude Oil Exports, Monthly**, which has a Series ID '**TOTAL.COEXPUS.M**'.

```
In [3]: # Convert to pandas dataframe
        eia_data = pd.DataFrame(eia_api.data_by_series(series='TOTAL.COEXPUS.M'));
```

### 4.1.2  Data Cleaning

Taking a look at the dataframe, we can observe some evident inconsistencies.

Firstly, the dataframe provided by the EIA is not of the standard format **datetime**, which pandas indexing supports and provides extensive facility to. We would be writing a function which makes the index a **datetime** object so that we can convert the dataframe to a **datetime**-index dataframe for more compatibility with pandas, hmms, and pgmpy.

```
In [4]: import datetime # Using the datetime library

        def convert_to_datetime(input):
                return datetime.datetime.strptime(input[:9], "%Y %m ").date();

        # Apply to entire index
        eia_data.index = eia_data.index.map(convert_to_datetime);
        # Convert dataframe index to datetime64[ns] index
        eia_data.index = pd.to_datetime(eia_data.index);
        # pgmpy stores the column names as the variable name
        eia_data.columns = ['TOTAL.COEXPUS.M'];
```

The second issue are **holes** in the data i.e. rows marked by a '-' (a single dash). We would be replacing these dashes by **np.nan** so that we can use pandas to fill in the holes. Usually the prevalance of these holes is very rare, but just to be on the safe side to ensure we can possibly download every dataset.

```
In [5]:  # Replace the '-' with np.nan
         eia_data.replace('-', np.nan, regex=True, inplace=True);
         # Backward fill the holes, by filling them with the data infront.
         eia_data.fillna(method='bfill', inplace=True);
```

Together, we can create a function carrying out the entire process so that we can easily clean EIA data in one step.

```
In [6]:  def clean_EIA(data):
             data.replace('-', np.nan, regex=True, inplace=True);
             data.fillna(method='bfill', inplace=True);

             data.index = data.index.map(convert_to_datetime);
             data.index = pd.to_datetime(data.index);
```

The dataframe is now a time series dataframe which could be plotted as a time-series dataframe.

```
In [7]:  import matplotlib.pyplot as plt

         fig, ax = plt.subplots(figsize=(20,6));
         ax.plot(eia_data);
```

Now, we shall be taking a look at the FRED data. Similar to EIA-python, the fredapi requires us to register with FRED API so that we can access data. We would download the **Spot Crude Oil Price: West Texas Intermediate**, having the Series ID '**WTISPLC**.'

```
In [8]:  from fredapi import Fred

         # FRED API key
         fred_key = "XX";

         # Initiates a session with the FRED datacenter to recieve datasets
         fred = Fred(api_key=fred_key);

         # Retrieve data from FRED API
         fred_data = pd.DataFrame(fred.get_series('WTISPLC'), columns=['WTISPLC']);
```

It is evident that the FRED, though still being a government organization, has 'ready-to-use' / 'plug'n play' data of useable quality compared to the EIA. Fortunely, we will not be having to clean data obtained from the FRED.

### 4.1.3 Constructing the training, validation and testing datasets (Data transformation)

As mentioned in the thesis, we need to divide our data in three portions: the training dataset, the validation dataset, and the training dataset. Given that we would be using a number of datasets from the FRED and the EIA, we would have to amalgamate these datasets into one dataframe and then slice the dataframe accordingly.

The train, validation, and test datasets are to be observed with a ratio of 80 : 10 : 10, which is a popular ettiquette

The choice of datasets has been described in the thesis.

```
In [9]:  # Dataset series ID from the EIA

         datasets_eia = [

                                 'STEO.RGDPQ_NONOECD.M',
                                 'STEO.RGDPQ_OECD.M',

                                 'STEO.PAPR_NONOPEC.M',
                                 'STEO.PAPR_OPEC.M',

                                 'STEO.PATC_OECD.M',
                                 'STEO.PATC_NON_OECD.M',

                                 'STEO.COPRPUS.M',
                                 'STEO.CORIPUS.M',
                                 'PET.MCRIMXX2.M',

                                 'STEO.FOREX_WORLD.M',

                                 'STEO.PASC_OECD_T3.M',

                                 'STEO.COPS_OPEC.M',
                                 'STEO.COPC_OPEC.M',

                                 'STEO.T3_STCHANGE_OOECD.M',
                                 'STEO.T3_STCHANGE_NOECD.M',
                         ];

         # Dataset series ID from the FRED

         datasets_fred = [
                                 'CPIENGSL',
                                 'CAPG211S',
                                 'CAPUTLG211S',
                                 'IPG211S',
                                 'IPG211111CN',
                                 'INDPRO',
                                 'IPN213111N',
                                 'PCU211211',

                         ];
```

To construct the training, validation, and testing datasets, we need to first **concatenate** the datasets into one dataframe, and then slice it.

```
In [10]:  data_merge = []; # List of dataframes to be concatenated
```

```
# Adding EIA datasets

for series_id in datasets_eia:
        df = pd.DataFrame(eia_api.data_by_series(series=series_id));
        clean_EIA(df);
        df.columns = [series_id];
        data_merge.append(df);

# Adding FRED datasets

for series_id in datasets_fred:
        df = pd.DataFrame(fred.get_series(series_id), columns=[series_id]);
        data_merge.append(df);
```

We have to create two additional columns; one which has the current crude oil price, and the other for the price of crude oil next month (forecast). This will be used to forecast the price of oil and hence allow us to make buy/sell decisions based on that forecast.

```
In [11]: datasets = datasets_eia + datasets_fred + ['WTISPLC', 'forecast'];

         current  = pd.DataFrame(fred.get_series('WTISPLC'), columns=['WTISPLC']);
         forecast = pd.DataFrame(fred.get_series('WTISPLC').shift(-1),
                                 columns=['forecast']);

         data_merge.append(current);
         data_merge.append(forecast);
```

We have to amalgamate all datasets together in a single dataframe, therefore we would use the pandas **concatenate** function. This would allow us to find the intersection of the date intervals of all dataframes and construct a single dataframe on a common time interval.

```
In [12]: data = pd.concat(data_merge, axis=1, join='inner');
```

Slicing our dataframe in train, validation, and testing datasets,

```
In [13]: train_data = data[: int(data.shape[0] * 0.80)];
         vald_data = data[int(0.80 * data.shape[0]) : int(0.90 * data.shape[0])];
         test_data = data[int(0.90* data.shape[0]) : int(data.shape[0])];
```

### 4.1.4   Data Discretisation

The data we have collected is non-categorical data; it is unlabelled and continuous. Belief networks have variables, each having discrete **states**, and therefore we have to reduce our data from prices to a set of states, such as bull, bear, and stagnant markets. In order to detect these (hidden) states, we have to use graphical models called **Hidden Markov Models**. The process of detecting hidden states in time-series data is called **Regime Detection**.

## 4.2 Regime Detection

We would be using Python library called hmms for implementing the **Hidden Markov Models**.

A **Hidden Markov Model** is a 5-tuple $(Q, \sum, \Pi, A, B)$, where $Q = \{q_1, \cdots, q_N\}$ is a finite set of $\mathcal{N}$ states, $\sum = \{s_1, \cdots, s_N\}$ is the set of $\mathcal{M}$ possible symbols (emissions) in the language, $\Pi = \{\pi_i\}$ is the initial probability vector, $A = \{a_{ij}\}$ is the state transition probability matrix, and $B = \{b_i(v_k)\}$ is the emission probability matrix. The HMM can be denoted by $\lambda = (\Pi, A, B)$.

For detecting regimes in time-series data, we would be using **Hidden Markov Models**, with the difference between consecutive months being the symbols $\sum$ (1 - increase / 0 - decrease), the hidden states, $Q$ being the **bull**, **bear**, **stagnant** market regimes.

Let us use '**WTISPL**' (Spot Crude Oil Price: West Texas Intermediate) of and try to identify regimes in the time series.

```
In [14]: import hmms
```

```
In [15]: price = train_data['WTISPLC'];
```

We will now try to transform the time series which represents the output emissions, with 1 representing an increase in the price from the previous month and 0 representing a decrease in the price of the oil.

```
In [16]: # The first value is NaN as there is not a previous month to compare with
         price_diff = price.diff()[1:];

         # Replacing the change with 1 if positive, else 0
         e_seq = np.array(price_diff.apply(lambda x: 1 if x > 0 else 0).values);
```

Given that we have obtained the output (observed emission sequence), we can now use the **Baum-Welch algorithm** to learn the parameters of the HMM generating this data.

We have earlier described the **Baum-Welch algorithm** in the Background and Literature review, and we would be using the implementation provided hmms to learn the parameters.

**IT IS VERY IMPORTANT** to note we can **only** use the training data to train the HMM as we are assuming to be blind to the testing data. However, we would be observing predictions on the validation dataset and will tune our model to fit it, and we would be using the testing dataframe to test the final performance of the tuned model after validation.

We will create a model with random parameters, that will be eventually trained to match the data - a discrete time HMM of three hidden states (bull, bear, or stagnant) and two output variables (increase or decrease).

```
In [17]: dhmm_r = hmms.DtHMM.random(3 , 2);
```

Given that the `hmms.DtHMM` takes a list of arrays no creater than length 32, we will have to split our array in arrays each of length 32 or less.

```
In [18]: e_seq = np.array_split(e_seq, 32);
```

### 4.2.1 Baum-Welch Algorithm

We would now be using the **Baum-Welch algorithm** to learn the parameters of the HMM generating the time-series.

The probability of the reestimated model after each iteration should ideally be closer that the (unknown) generator's model, however chances might be the estimation fell in the local optima.

Unfortunately, the financial time-series data do not have fixed parameters, so the HMM has to be trained everytime when live-trading, when using the **k-fold cross-validation** training method.

```
In [19]: dhmm_r.baum_welch(e_seq, 100); # 100 iterations
```

We have now learnt the parameters generating the emission sequence.

```
In [20]: hmms.print_parameters( dhmm_r );
```

Initial probabilities () :

```
          0
0  0.209758
1  0.650148
2  0.140094
```

Transition probabilities matrix (A):

```
          0         1         2
0  0.383857  0.550736  0.065407
1  0.056177  0.435758  0.508065
2  0.488859  0.000035  0.511106
```

Emission probabilities matrix (B):

```
          0         1
0  0.999584  0.000416
1  0.000149  0.999851
2  0.286833  0.713167
```

### 4.2.2 Viterbi Algorithm

Now, given we now have parameters $\lambda$ and the emitted observation sequence e_seq, we can use the **Viterbi Algorithm** to identify the most likely **state-transition** path (i.e. **market regimes**) in the financial time-series.

```
In [21]: ( log_prob, s_seq ) = dhmm_r.viterbi(np.concatenate(e_seq).ravel());
```

### 4.2.3 Multicolored time series plot

Now, we will be plotting this graph in a **multicolored** time-series plot to observe how well the regimes have been identified.

First, we will have to make a dataframe which has the both the price and the associated regime the time-series data is in.

```
In [22]: # Add price
         price_plot = pd.DataFrame(price[1:], index=price[1:].index);

         # Add a column representing the regime
         price_plot['Regime'] = s_seq;

         # Add a column representing the increase or decrease in price
         price_plot['diff'] = price_diff;
```

We do not know, however, which state represents which regime. Given that the bull regimes should have a high positive change in price, bear regimes should have a high negative change, and stagnant regimes are closer to zero, we can use these properties to tell which state represents which regime.

```
In [23]: # Get means of all assigned states
         means = price_plot.groupby(['Regime'])['diff'].mean();
         lst_1 = means.index.tolist();
         lst_2 = means.sort_values().index.tolist();

         map_regimes = dict(zip(lst_2, lst_1));

         price_plot['Regime'] = price_plot['Regime'].map(map_regimes);
```

Plotting the data as **multi-colored** time series:

```
In [24]: import matplotlib.dates as mdates
         import matplotlib.patches as mpatches
         from matplotlib.collections import LineCollection
         from matplotlib.colors import Colormap, ListedColormap, BoundaryNorm

         fig, ax1 = plt.subplots(figsize=(20,8));
         ax.plot(price_plot['WTISPLC']);

         # Make 0 (Bear) - red, 1 (Stagnant) - blue, 2 (Bull) - green

         cmap    = ListedColormap(['r','b','g'],'indexed');
         norm    = BoundaryNorm(range(3 + 1), cmap.N);
         inxval  = mdates.date2num(price_plot['WTISPLC'].index.to_pydatetime());
         points  = np.array([inxval, price_plot['WTISPLC']]).T.reshape(-1, 1, 2);
         segments = np.concatenate([points[:-1], points[1:]], axis=1);

         lc = LineCollection(segments, cmap=cmap, norm=norm);
```

```
lc.set_array(price_plot['Regime']);
plt.gca().add_collection(lc);
plt.xlim(price_plot['WTISPLC'].index.min(), price_plot['WTISPLC'].index.max());
plt.ylim(price_plot['WTISPLC'].min(), price_plot['WTISPLC'].max());

r_patch = mpatches.Patch(color='red', label='Bear');
g_patch = mpatches.Patch(color='green', label='Bull');
b_patch = mpatches.Patch(color='blue', label='Stagnant');

plt.legend(handles=[r_patch, g_patch, b_patch]);

plt.show();
```

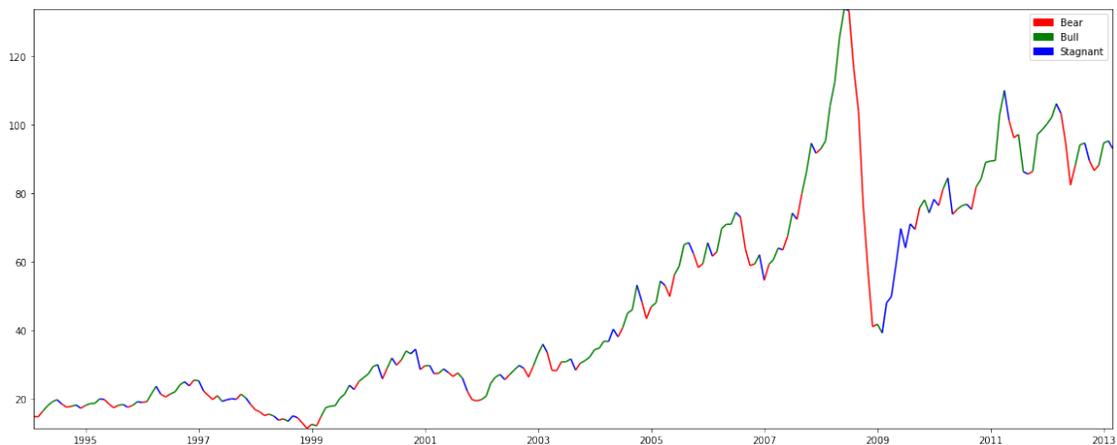

We can observe from the graph that there are periods of bear runs, bull runs, and stagnant periods. Now, we need to apply a similar method to all the training data so that we can discretise our data and use it as an input to our belief network.

### 4.2.4 Discretising dataframes using Hidden Markov Models

We will write a function that trains an HMM, identifies the sequence of hidden states, and then constructs a dataframe of all the variables and stores it as training dataset.

Now, we shall apply the same method to the entire training dataset so that we can discretise it.

First, we shall learn all the parameters of Hidden Markov Models of all the variables and store them.

```
In [25]: for series_id in datasets:
             if series_id == 'forecast':
                 break;
             else:
                 dhmm = hmms.DtHMM.random(3,2);
                 data_diff = train_data[series_id].diff()[1:];
                 emit_seq = np.array_split(data_diff.apply(
                         lambda x: 1 if x > 0 else 0).values, 32);
```

```
        dhmm.baum_welch(emit_seq, 100);
        path = "./hmms/" + series_id.replace(".", "_");
        dhmm.save_params(path);
```

Now, we shall be constructing the discretised training dataframe.

```
In [26]: disc_test = pd.DataFrame(index = train_data[1:].index);

        for series_id in datasets:
            path = "./hmms/" + series_id.replace(".", "_") + ".npz";
            if series_id == 'forecast':
                dhmm = hmms.DtHMM.from_file('./hmms/WTISPLC.npz');
            else:
                dhmm = hmms.DtHMM.from_file(path);
            data_diff =  train_data[series_id].diff()[1:];
            emit_seq = np.array(data_diff.apply(
                        lambda x: 1 if x > 0 else 0).values);
            ( log_prob, s_seq ) =  dhmm.viterbi(emit_seq);
            disc_test[series_id] = s_seq;

        disc_test.to_csv("./data/train_data.csv"); # Saving to CSV
```

### 4.2.5 (Recommended) Plotting Regime Switch plots

In order to verify if the **HMM** has correctly identified regimes in all datasets, we should plot the **regime-switching models** of all datasets and observe if the **bull**, **bear**, and **stagnant** states have been correctly identified. The graphical representation allows us to understand how the regimes have been learnt.

Omission of this step can result in the **HMM** being incorrectly trained, hence identifying incorrect regimes consequently affecting the belief network's training and prediction process.

```
In [27]: states = pd.read_csv("./data/train_data.csv", index_col=0);

        for series_id in datasets:

            df = pd.DataFrame(index=train_data[1:].index);
            df[series_id] = train_data[series_id][1:];
            df['Diff'] = train_data[series_id].diff()[1:];
            df['Regime'] = states[series_id];

            # Get means of all assigned states
            means = df.groupby(['Regime'])['Diff'].mean();
            lst_1 = means.index.tolist();
            lst_2 = means.sort_values().index.tolist();

            map_regimes = dict(zip(lst_2, lst_1));
            df['Regime'] = df['Regime'].map(map_regimes);
```

```
cmap     = ListedColormap(['r','b','g'],'indexed');
norm     = BoundaryNorm(range(3 + 1), cmap.N);
inxval   = mdates.date2num(df[series_id].index.to_pydatetime());
points   = np.array([inxval, df[series_id]]).T.reshape(-1, 1, 2);
segments = np.concatenate([points[:-1], points[1:]], axis=1);

lc = LineCollection(segments, cmap=cmap, norm=norm);
lc.set_array(df['Regime']);
plt.gca().add_collection(lc);
plt.xlim(df[series_id].index.min(), df[series_id].index.max());
plt.ylim(df[series_id].min(), df[series_id].max());

r_patch = mpatches.Patch(color='red', label='Bear');
g_patch = mpatches.Patch(color='green', label='Bull');
b_patch = mpatches.Patch(color='blue', label='Stagnant');

plt.legend(handles=[r_patch, g_patch, b_patch]);

name = "./plots/" + series_id.replace(".", "_") + ".png";

plt.savefig(name);
plt.close();
```

We have sucessfully discretised our training dataset and would now be using it to train the Belief Network.

### 4.2.6 Learning Bayesian Network using Hill Climbing

Given that we have discretised our training dataset, we can now use pgmpy to construct a belief network of all the variables.

We would be using the **Hill Climbing** approach to learn the belief network. We have given a brief description of the Hill Climbing algorithm in the thesis and would now be using its implementation pgmpy to learn the structure of the oil markets.

We shall begin with importing the relevant modules from pgmpy in Python.

Given we would be using the *BIC Scoring Algorithm* as the scoring function for Hill Climbing, we will import the relevant functions from the relevant module in pgmpy.

```
In [28]: from pgmpy.models import BayesianModel
         from pgmpy.estimators import HillClimbSearch
         from pgmpy.estimators import BayesianEstimator
         from pgmpy.estimators import BicScore, K2Score, BdeuScore

         # Retrieve training set
         train_data = pd.read_csv("./data/train_data.csv", index_col=0);
```

We shall now be performing a Hill-Climbing search. As difficult as it seems, it is a quite straightforward process.

We construct a instance of a Bayesian model, having the initial structure constructed by expert knowledge from EIA. We shall then use Hill Climbing to 'attach' remaining macroeconomic variables to the constructed model.

```python
In [29]: # Initialise Hill Climbing Estimator
         hc = HillClimbSearch(train_data, scoring_method=K2Score(train_data));
         expert = BayesianModel();
         expert.add_nodes_from(datasets);
         expert.add_edges_from([
                               ('STEO.PAPR_NONOPEC.M', 'WTISPLC'),
                               ('STEO.PAPR_OPEC.M', 'WTISPLC'),
                               ('STEO.PATC_OECD.M', 'WTISPLC'),
                               ('STEO.PATC_NON_OECD.M', 'WTISPLC'),
                               ('STEO.RGDPQ_OECD.M', 'STEO.PATC_OECD.M'),
                               ('STEO.RGDPQ_NONOECD.M', 'STEO.PATC_NON_OECD.M'),
                              ]);

         model = hc.estimate(expert); # Performs local hill climb search

         model.fit(train_data,
                   state_names=dict(map(lambda e: (e, [0, 1, 2]), datasets)),
                   estimator=BayesianEstimator, prior_type="K2");

In [30]: import networkx as nx
         import pylab as plt

         G=nx.Graph();
         G.add_edges_from(model.edges());
         pos = nx.spring_layout(G);
         nx.draw_networkx_nodes(G, pos, node_size = 10);
         nx.draw_networkx_edges(G, pos, arrows=True);
         plt.figure(5,figsize=(20,10));
```

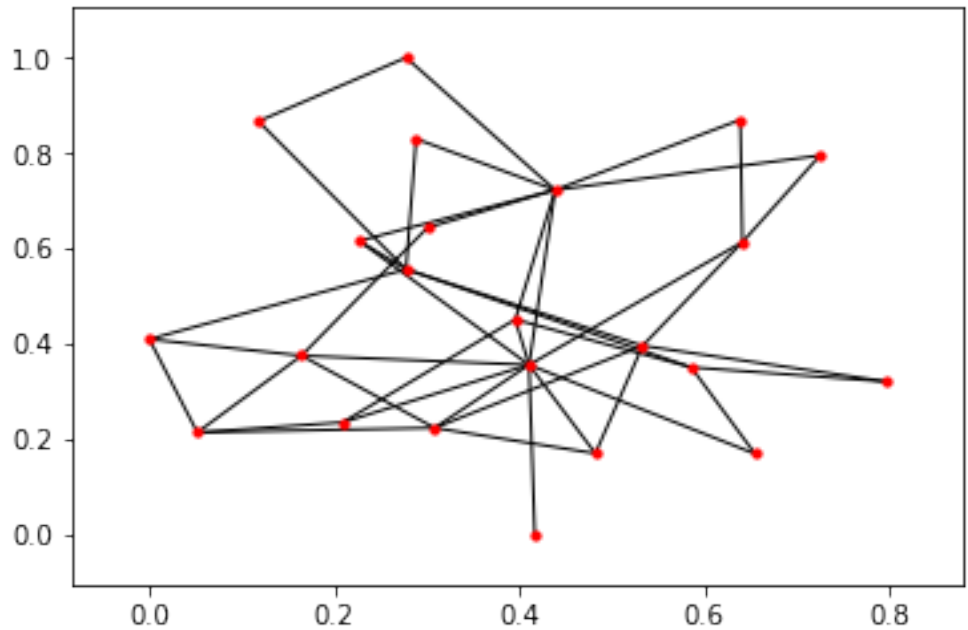

```
<matplotlib.figure.Figure at 0x119b316a0>
```

We can also represent our model in the form of a directed graph, using NetworkX.

We have now fitted our model using the Hill Climb search, and now can make inferences using forecasts as evidence.

## 4.3   Validation

We have successfully fitted our model to the data and need to test our model now.

We will now be using our **validation** dataset and would be making predictions on that and would be accordingly adjusting our model if we are not satisfied by the performance.

We first have to discretise the validation dataset. We would be using the HMMs trained on the **training set**. We **CANNOT** use the **validation set** to train anything, including HMMs!

```
In [31]: discrete_vald = pd.DataFrame(index = vald_data[1:].index);

         for series_id in datasets:
             path = "./hmms/" + series_id.replace(".", "_") + ".npz";
             if series_id == 'forecast':
                 dhmm = hmms.DtHMM.from_file('./hmms/WTISPLC.npz');
             else:
                 dhmm = hmms.DtHMM.from_file(path);
             data_diff =  vald_data[series_id].diff()[1:];
             emit_seq = np.array(data_diff.apply(lambda x: 1 if x > 0 else 0).values);
```

```
        ( log_prob, s_seq ) = dhmm.viterbi(emit_seq);
        discrete_vald[series_id] = s_seq;

    discrete_vald.to_csv("./data/validation_data.csv"); # Saving to CSV
```

Now, we shall be plotting this data to see how well the trained HMMs predicted regimes on the valdiation dataset.

```
In [32]: states = pd.read_csv("./data/validation_data.csv", index_col=0);

    for series_id in datasets:

        df = pd.DataFrame(index=vald_data[1:].index);
        df[series_id] = vald_data[series_id][1:];
        df['Diff'] = vald_data[series_id].diff()[1:];
        df['Regime'] = states[series_id];

        # Get means of all assigned states

        means = df.groupby(['Regime'])['Diff'].mean();

        lst_1 = means.index.tolist();
        lst_2 = means.sort_values().index.tolist();

        map_regimes = dict(zip(lst_2, lst_1));
        df['Regime'] = df['Regime'].map(map_regimes);

        cmap   = ListedColormap(['r','b','g'],'indexed');
        norm   = BoundaryNorm(range(3 + 1), cmap.N);
        inxval = mdates.date2num(df[series_id].index.to_pydatetime());
        points = np.array([inxval, df[series_id]]).T.reshape(-1, 1, 2);
        segments = np.concatenate([points[:-1], points[1:]], axis=1);

        lc = LineCollection(segments, cmap=cmap, norm=norm);
        lc.set_array(df['Regime']);
        plt.gca().add_collection(lc);
        plt.xlim(df[series_id].index.min(), df[series_id].index.max());
        plt.ylim(df[series_id].min(), df[series_id].max());

        r_patch = mpatches.Patch(color='red', label='Bear');
        g_patch = mpatches.Patch(color='green', label='Bull');
        b_patch = mpatches.Patch(color='blue', label='Stagnant');

        plt.legend(handles=[r_patch, g_patch, b_patch]);

        name = "./plots/" + series_id.replace(".", "_") + "_VALIDATION.png";
```

```
            plt.savefig(name);
            plt.close();
```

Now, we would be predicting the validation sets (forecast for Spot Crude Oil price, WTI Monthly).

For that, we have to drop the forecast column and then do an inference on the model.

```
In [33]: # Record real data observation, to be compared with the predicted one
         vald_real = states['WTISPLC'].as_matrix();

         # Drop the real data observation so that it does not bias prediction
         vald_data_new = states.drop('forecast', axis=1);

         # Inference on the constructed graphical model
         vald_prediction = model.predict(vald_data_new);

         # Retrieve it as an array so we can compare with real value
         pred_value_vald = vald_prediction['forecast'].as_matrix();

In [34]: print("\nPredicted Value: ");
         print(pred_value_vald);
         print("\nReal Value: ");
         print(vald_real);

         error = np.mean(vald_real != np.roll(pred_value_vald, 1));
         #error = np.mean(vald_real != pred_value_vald);
         print("\nError: ");
         print(error * 100);
```

```
Predicted Value:
[2 0 2 1 1 1 1 0 1 0 2 0 2 2 1 1 1 1 1 1 1 0 1 0 2 2 1 1]

Real Value:
[0 1 0 0 2 2 2 1 2 1 2 1 0 0 2 2 2 2 2 2 2 1 2 1 0 0 2 2]

Error:
67.85714285714286
```

Now, we are going to trade on the predictions and compare the performance of the algorithm on returns, beginning with one oil share.

If we have achieved reasonable satisfaction of results on the **validation set**, we can now move to using the **test** set, and check the performance of our algorithm on that.

The **validation step** is to adjust the model if the error is too high. In this case, we can start again by learning the Bayesian model via the Hill Climbing method and observe the change in performance. We should however, **not** use the **test set** unless and until we are satisfied with the performance of the model on the **validation set**.

One way of assessing the quality of a network structure is by examining the **connectedness** of the graph, ensuring there are almost no forests, and **the variables which are being forecasted** are part of a denser tree. If we happen to see disconnected trees, we should run our Hill Climbing method again to ensure that the Hill Climber **converges** to either ideally a **global maximum** or atleast a better **local maximum**.

### 4.3.1 Testing

Similar to what we did with the validation data set, we first have to discretise the test dataset, and predict it on the *BayesianModel*.

```
In [35]: discrete_test = pd.DataFrame(index = test_data[1:].index);

         for series_id in datasets:
             path = "./hmms/" + series_id.replace(".", "_") + ".npz";
             if series_id == 'forecast':
                 dhmm = hmms.DtHMM.from_file('./hmms/WTISPLC.npz');
             else:
                 dhmm = hmms.DtHMM.from_file(path);
             data_diff = test_data[series_id].diff()[1:];
             emit_seq = np.array(data_diff.apply(lambda x: 1 if x > 0 else 0).values);
             ( log_prob, s_seq ) = dhmm.viterbi(emit_seq);
             discrete_test[series_id] = s_seq;

         discrete_test.to_csv("./data/test_data.csv");
```

We would now import the (discretised) test dataframe, remove the column containing forecast column as we are predicting it, and input it in our learnt model. We would compare the output to the real values and make a final conclusion of the reliability of the model.

```
In [36]: discrete_test = pd.read_csv("./data/test_data.csv", index_col=0);

         # Record real data observation, to be compared with the predicted one
         test_real = discrete_test['WTISPLC'].as_matrix();

         # Drop the real data observation so that it does not bias prediction
         test_data_new = discrete_test.drop('forecast', axis=1);

         # Inference on the constructed graphical model
         test_prediction = model.predict(test_data_new);

         # Retrieve it as an array so we can compare with real value
         pred_value_test = test_prediction['forecast'].as_matrix();

In [37]: print("\nPredicted Value: ");

         # This is the price, not the forecast
         print(pred_value_test);
         print("\nReal Value: ");
```

```python
print(test_real);

# Shift to get forecast
error = np.mean(test_real != np.roll(pred_value_test, 1));
#error = np.mean(test_real != pred_value_test); # Shift to get forecast
print("\nError: ");
print(error * 100);
```

```
Predicted Value:
[2 1 1 1 2 0 2 0 2 1 0 2 2 1 0 2 2 1 0 1 1 0 2 0 2 0 1 0]

Real Value:
[0 2 2 2 1 0 1 0 2 1 0 0 2 1 0 0 2 1 2 2 1 0 1 0 1 0 0]

Error:
42.857142857142854
```

Now, we can use the discretised price predictions to construct a simple trading algorithm, which takes '0' as a shorting signal, '2' for a long position, and '1' for no action. We assume we start with one barrel of oil and we would be trading it to acquire a larger unit.

```python
In [39]: test_price = pd.DataFrame(test_data['WTISPLC'], columns=['WTISPLC']);
         test_signal = pd.DataFrame(test_prediction, columns=['forecast']);
         test_sheet = pd.concat([test_price, test_signal], axis=1, join='inner');

         trades = [test_sheet['WTISPLC'].iloc[0]];

         position = False; # True for Long, False for Short

         for i in range(len(test_sheet)-1):
             if test_sheet['forecast'].iloc[i+1] == 0:
                 trades.append(trades[-1]);
             elif test_sheet['forecast'].iloc[i+1] == 2:
                 if position == False:
                     position = True;
                     trades.append(trades[-1]);
                 else:
                     trades.append(test_sheet['WTISPLC'].iloc[i+1]);
             else:
                 if position == False: # If Short, remains same price
                     trades.append(trades[-1]);
                 else: # If Long, price inc
                     trades.append(test_sheet['WTISPLC'].iloc[i+1]);

         eia_forecast = pd.read_csv("./data/eia_forecast.csv", index_col=0);
         eia_forecast.index = test_signal.index;
```

```
test_performance = pd.DataFrame(trades, index = test_signal.index,
                                          columns=['performance']);
test_sheet = pd.concat([test_sheet, test_performance], axis=1, join='inner');

plt.plot(test_sheet['WTISPLC'], 'r');
plt.plot(test_sheet['performance'], 'g');
plt.plot(eia_forecast['eia_forecast'], 'b');

r_patch = mpatches.Patch(color='red', label='WTI Crude Oil Price');
g_patch = mpatches.Patch(color='green', label='Bayesian Model');
b_patch = mpatches.Patch(color='blue', label='EIA Forecast');

plt.legend(handles=[r_patch, g_patch, b_patch], loc = 'lower right');
```

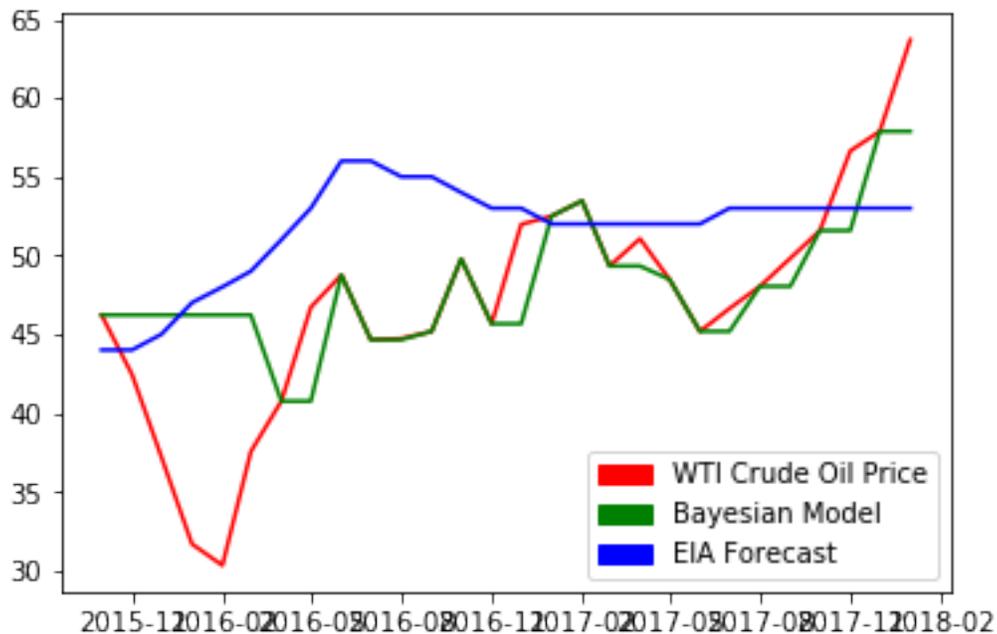

As we can see, the algorithm has successfully managed to hedge the risk in the early quarter of 2015, but did end up shorting in a bull run in late 2017/early 2018.

We can also compare the performance of the model with the performance of the EIA forecasts, where we can see the EIA forecasted a bull run when it was a bear run in early 2016. We can also similarly compare the performance in 2017 where both the EIA **Short Term Energy Outlook** (STEO) and the Bayesian Model predicted the the price will remain **stagnant**, however, in reality, it was a bull run.

Further research is necessary to statisically conclude with confidence if the model performs better than the EIA's STEO under different stress-events, which might be in turn, trading opportunities for oil traders.

# Chapter 5

# Conclusions & Future Work

The contribution of this thesis to science is to address scientific challenges of conducting forecasting the price of oil in real world operational settings. The proposed Computational Finance and Machine Learning methodologies are highly applicable in a number of real world settings. Firstly, they analyse the structure of the oil markets by constructing a graphical model associating different macroeconomic and physical market factors *without* any expert assistance, thus enabling speculators, risk managers, and energy policy makers to have a greater understanding of the structure of the oil markets. Secondly, it analyses the risk in the energy markets by providing forecasts using current economic situation as evidence. Thirdly, it provides an automated trading mechanism which learns and improves its trading decisions as time passes, utlimately resulting in a higher alpha for a commodities trader.

In conclusion, this dissertation contributes to existing literature in a number of ways. Firstly, it contributes to the original research of replacing EGARCH-M derived views with Bayesian Model derived views for the Black-Litterman model [15]. Secondly, the idea of using time-series data discretised by Hidden Markov Models as inputs to Belief Networks is novel. Thirdly, the dissertation presents a working trading mechanism which is directly deployable in the commodity markets and is entirely capable of independent decision making. Secondly, given the structural learning techniques employed, it is almost an autonomous decision making system which requires absolutely no prior expert knowledge, other than the selection of datasets which is carried out by commodity market analysts. Fourthly, the idea of using time-series data discretised by Hidden Markov Models as inputs to Belief Networks is novel. Fifthly, the level of abstraction for graphical models provided by the Python mod-ules allowed us to understand the design process in theoretical context. Sixthly, it provides forecasts using a systematic, event-driven, global macro strategy which takes in account mega geopolitical and macroeconomic changes, thus is capable of generating a higher return than funds based on a high-frequency or fixed-income strategy. Seventhly, it allows us to construct better models of energy markets, hence allowing energy policy makers to understand the underlying structure of the oil markets and use it to respond to different economic events by drafting more effective and sound policy. Lastly, this research is amalgamating a multitude of existing research and experimentation in multiple disciplines and applying it in the commodity markets, in an effort of increasing the alpha for quantitative commodity traders.

## 5.1 Future Work

Though this research successfully pursued objectives with predefined scopes, it also inspired future research research direction based on existing work. A number of subsequent topics may worth further investigation as a continuation of this research. Firstly, we could use explore new and more optimized Structure learning algorithms such as max-min hill climbing algorithm which would enhance the structure learning process by scaling up the datasets involving hundreds of variables [19]. Secondly, future research could go beyond using existing probabilisitic graphical models. New graphical models could be constructed either entirely on a new foundation or by combining features of existing graphical models. There are a number of new probabilistic graphical models have been constructed for algorithmic trading, such as Gated Bayesian Networks [17]. Thirdly, we could incorporate concepts from reinforcement learning in order to construct better models of the commodities markets by modelling returns as reward to actions [51]. Fourthly, we could apply concepts from Game Theory in order to model relationships between OPEC and non-OPEC producers, in an attempt to understand OPECs behaviour by trying to maintain supply at a certain level to maintain an oil price [23]. Finally, we could be using different market strategies such as arbitrage, hedging, and pairs trading in order for shorter-term trading.

# Bibliography

[1] Ankur Ankan. *Learning Bayesian Networks*. 2017. URL: https://github.com/pgmpy/pgmpy_notebook/blob/master/notebooks/9.%20Learning%20Bayesian%20Networks%20from%20Data.ipynb.

[2] Ankur Ankan. *pgmpy: Probabilistic Graphical Models using Python*. English. Proc. of the 14th Python in Science Conf. (SciPy 2015). 11 pp.

[3] Ankur Ankan and Abinash Panda. *pgmpy*. https://github.com/pgmpy. 2015.

[4] Alejandro Badel and Joseph McGillicuddy. *Oil Prices and Inflation Expectations: Is There a Link?* Tech. rep. Federal Reserve Bank, St. Louis, 2015, p. 13. URL: https://www.stlouisfed.org/~/media/Publications/Regional-Economist/2015/July/Oil.pdf.

[5] D. Barber. *Bayesian Reasoning and Machine Learning*. 04-2011. In press. Cambridge University Press, 2011. URL: http://www.cs.ucl.ac.uk/staff/d.barber/brml.

[6] D. Barber, A.T. Cemgil, and S. Chiappa. *Bayesian Time Series Models*. Bayesian Time Series Models. Cambridge University Press, 2011. ISBN: 9780521196765. URL: https://books.google.com.pk/books?id=k4z6mOFsEv8C.

[7] Christian Baumeister, Lutz Kilian, and Thomas K. Lee. *Are there Gains from Pooling Real Time Oil Price Forecasts?* Tech. rep. Ann Arbor, MI 48109-1220, USA: University of Michigan, Department of Economics, 2014, p. 24. URL: https://www.eia.gov/workingpapers/pdf/oilprice_forecasts.pdf.

[8] Christiane Baumeister Baumeister, Pierre Guérin, and Lutz Kilian. *Do High-Frequency Financial Data Help Forecast Oil Prices? The MIDAS Touch at Work*. Tech. rep. 2014-11. Bank of Canada, 2014, p. 34. URL: https://www.banqueducanada.ca/wp-content/uploads/2014/03/wp2014-11.pdf.

[9] Bruce Bawks. *What drives crude oil prices?* Tech. rep. U.S. Energy Information Administration, 2018, p. 23. URL: https://www.eia.gov/finance/markets/crudeoil/reports_presentations/crude.pdf.

[10] Bruce Bawks. *What drives crude oil prices: Demand Non-OECD*. URL: https://www.eia.gov/finance/markets/crudeoil/demand-nonoecd.php.

[11] Bruce Bawks. *What drives crude oil prices: Demand OECD*. URL: https://www.eia.gov/finance/markets/crudeoil/demand-oecd.php.

[12]   Bruce Bawks. *What drives crude oil prices: Overview.* URL: `https://www.eia.gov/finance/markets/crudeoil/`.

[13]   Bruce Bawks. *What drives crude oil prices: Supply Non-OPEC.* URL: `https://www.eia.gov/finance/markets/crudeoil/supply-nonopec.php`.

[14]   Bruce Bawks. *What drives crude oil prices: Supply OPEC.* URL: `https://www.eia.gov/finance/markets/crudeoil/supply-opec.php`.

[15]   Steven L. Beach and Alexei G. Orlov. "An application of the Black–Litterman model with EGARCH-M-derived views for international portfolio management". In: *Financial Markets and Portfolio Management* 21.2 (2007). DOI: `10.1007/s11408-007-0046-6`. URL: `https://doi.org/10.1007/s11408-007-0046-6`.

[16]   Joscha Beckmann, Vipin Arora, and Robert Czudaj. *The Relationship Between Oil Prices and Exchange Rates: Theory and Evidence.* University of Bochum and Kiel Institute for the World Economy. U.S. Energy Information Administration. URL: `https://www.eia.gov/finance/markets/reports_presentations/2017/beckmann.pdf`.

[17]   Marcus Bendtsen. *Gated Bayesian networks.* Vol. 1851. Linköping University Electronic Press, 2017.

[18]   Havar Blakset. *OPEC Production and Consequences for short term Oil Price.* EIA Workshop on Financial & Physical Oil Market Linkages. Rystad Energy. Sept. 19, 2017. URL: `https://www.eia.gov/finance/markets/reports_presentations/2017/blakset.pdf` (visited on 04/10/2018).

[19]   Laura E Brown, Ioannis Tsamardinos, and Constantin F Aliferis. "A novel algorithm for scalable and accurate Bayesian network learning". In: *Studies in health technology and informatics.* 2004.

[20]   Bawks Bruce. *An analysis of 7 factors that influence oil markets with chart data updated monthly and quarterly.* Independent Statistics & Analysis. U.S. Energy Information Administration. URL: `https://www.eia.gov/finance/markets/crudeoil/reports_presentations/crude.pdf`.

[21]   Alexandra M Carvalho. "Scoring functions for learning Bayesian networks". In: *Inesc-id Tec. Rep* 12 (2009).

[22]   Robert Castelo and Arno Siebes. "Priors on network structures. Biasing the search for Bayesian networks". In: *International Journal of Approximate Reasoning* 24.1 (2000), pp. 39–57.

[23]   Yuwen Chang et al. "Oil supply between OPEC and non-OPEC based on game theory". In: 45 (Oct. 2014).

[24]   R.G. Cowell et al. *Probabilistic Networks and Expert Systems.* Jan. 2001.

[25] Nir Friedman Daphne Koller. *Probabilistic Graphical Models: Principles and Techniques*. 1st ed. Adaptive Computation and Machine Learning series. The MIT Press, 2009. ISBN: 0262013193,9780262013192. URL: `http://gen.lib.rus.ec/book/index.php?md5=8ac4fc1b72fdad0ec00029fb520bfce4`.

[26] Rakesh Dugad and UDAY B Desai. "A tutorial on hidden Markov models". In: *Signal Processing and Artifical Neural Networks Laboratory, Dept of Electrical Engineering, Indian Institute of Technology, Bombay Technical Report No.: SPANN-96.1* (1996).

[27] Wouter Duivesteijn. *Markov Chains and Hidden Markov Models*. 2006. URL: `http://www.cs.uu.nl/docs/vakken/cb/slides/slides3.pdf`.

[28] J Peter Ferderer. "Oil price volatility and the macroeconomy". In: *Journal of macroeconomics* 18.1 (1996), pp. 1–26.

[29] R. Dean Foreman. *Economic Activity & Oil Prices, OPEC Policies, Production and Price Responses*. EIA Workshop on Financial & Physical Oil Market Linkages. Saudi Aramco. Sept. 19, 2017. URL: `https://www.eia.gov/finance/markets/reports_presentations/2017/foreman.pdf` (visited on 04/10/2018).

[30] Jerome Friedman, Trevor Hastie, and Robert Tibshirani. *The elements of statistical learning*. Vol. 1. Springer series in statistics New York, 2001.

[31] Andrew D. Gordon, Aditya V. Henzinger Thomas A.; Nori, and Sriram K. Rajamani. *[ACM Press the - Hyderabad, India (2014.05.31-2014.06.07)] Proceedings of the on Future of Software Engineering - FOSE 2014 - Probabilistic programming*. 2014. ISBN: 9781450328654. DOI: `10.1145/2593882.2593900`. URL: `http://gen.lib.rus.ec/scimag/index.php?s=10.1145/2593882.2593900`.

[32] Jan J. J Groen. *(Again) Weaker Oil Prices: Demand,Supply, or Neither?* Research and Statistics Group, FRBNY. U.S. Energy Information Administration. URL: `https://www.eia.gov/finance/markets/reports_presentations/2016JanGroen.pdf`.

[33] James D Hamilton and Ana Maria Herrera. "Oil shocks and aggregate macroeconomic behavior: The role of monetary policy: A comment". In: *Journal of Money, Credit, and Banking* 36.2 (2004), pp. 265–286.

[34] F.A. Hayek. *The Fatal Conceit (Collected Works of Friedrich August Hayek)*, p. 77. ISBN: 0415008204,9780415008204. URL: `http://gen.lib.rus.ec/book/index.php?md5=CCBD1E1658FE2E5314D0E642E754ADE4`.

[35] Mortada Mehyar. *fredapi: Python API for FRED (Federal Reserve Economic Data)*. `https://github.com/mortada/fredapi`. 2017.

[36] mra1385. *EIA-python*. `https://github.com/mra1385/EIA-python`. 2015.

[37] Søren Højsgaard. *Graphical Models and Bayesian Networks*. 2016. URL: `http://people.math.aau.dk/~sorenh/misc/2016-gmbn-zurich/doc/bayesnet.pdf`.

[38]  Lutz Kilian. *Not All Oil Price Shocks Are Alike:Disentangling Demand and Supply Shocksin the Crude Oil Market.* University of Michigan and CEPR, 2008. URL: `http://www-personal.umich.edu/~lkilian/aer061308final.pdf`.

[39]  Daphne Koller and Nir Friedman. *Probabilistic Graphical Models: Principles and Techniques.* MIT Press, 2009.

[40]  Daphne Koller and Nir Friedman. "Probabilistic Graphical Models: principles and techniques". In: 2009. Chap. 18.3.4-18.3.6, p. 802.

[41]  *Lecture 1: Discrete Markov Chains.* 2015. URL: `http://www.mi.fu-berlin.de/wiki/pub/CompMolBio/MarkovKetten15/Lecture1.pdf`.

[42]  Chul-Yong Lee and Sung-Yoon Huh. "Forecasting Long-Term Crude Oil Prices Using a Bayesian Model with Informative Priors". In: *Sustainability* 9.12 (2017), p. 190. DOI: `10.3390/su9020190`. URL: `https://doi.org/10.3390/su9020190`.

[43]  Lukas Lopatovsky. *HMMs.* `https://github.com/lopatovsky/HMMs`. 2017.

[44]  Lukas Lopatovsky. *HMMs.* `https://github.com/lopatovsky/HMMs/blob/master/hmms.ipynb`. 2017.

[45]  Lukas Lopatovsky. "Lopatovsky, Lukas". MA thesis. Czech Technical University In Prague, 2017. URL: `https://dspace.cvut.cz/bitstream/handle/10467/69120/F8-DP-2017-Lopatovsky-Lukas-thesis.pdf`.

[46]  Brandon Malone. *Lecture Notes: Learning Bayesian Network Structures.* 2014. URL: `https://www.cs.helsinki.fi/u/bmmalone/probabilistic-models-spring-2014/StructureLearning.pdf`.

[47]  Dimitris Margaritis. *Learning Bayesian network model structure from data.* Tech. rep. Carnegie-Mellon University, School of Computer Science, Pittsburgh PA, 2003. URL: `https://www.cs.cmu.edu/~dmarg/Papers/PhD-Thesis-Margaritis.pdf`.

[48]  *Markov chains and Hidden Markov Models.* 2009. URL: `\url{http://www.inf.fu-berlin.de/lehre/WS05/aldabi/downloads/markov_part1.pdf}`.

[49]  Maxwell Maxwell. *Lecture 4: Introduction to pandas.* URL: `https://www.quantopian.com/lectures/introduction-to-pandas`.

[50]  R.E. Neapolitan. *Learning Bayesian Networks.* Artificial Intelligence. Pearson Prentice Hall, 2004. ISBN: 9780130125347.

[51]  Yuriy Nevmyvaka, Yi Feng, and Michael Kearns. "Reinforcement learning for optimized trade execution". In: *Proceedings of the 23rd international conference on Machine learning.* ACM, 2006, pp. 673–680.

[52]  *pandas.* `https://github.com/pandas-dev/pandas`. 2018.

[53]  Judea Pearl. *Probabilistic Reasoning in Intelligent Systems: Networks of Plausible Inference.* 1st ed. Morgan Kaufmann, 1988. ISBN: 1558604790,9781558604797. URL: `http://gen.lib.rus.ec/book/index.php?md5=9BDEE2434F2051324F7E71B76E150F22`.

[54] Jean-Philippe Pellet and André Elisseeff. "Using Markov blankets for causal structure learning". In: *Journal of Machine Learning Research* 9.Jul (2008), pp. 1295–1342.

[55] L. Rabiner and B. Juang. "An introduction to hidden Markov models". In: *IEEE ASSP Magazine* 3.1 (1986), pp. 4–16. ISSN: 0740-7467. DOI: `10.1109/MASSP.1986.1165342`.

[56] Riccardo Rebonato and Alexander Denev. *Portfolio Management Under Stress*. Cambridge University Press, 2009. DOI: `10.1017/cbo9781107256736`. URL: `https://doi.org/10.1017/cbo9781107256736`.

[57] Gallager R.G. *Stochastic Processes: Theory for Applications*. Cambridge University Press, 2013, p. 164. ISBN: 978-1-107-03975-9. URL: `http://gen.lib.rus.ec/book/index.php?md5=97e9ba867cf373396f2343c291de7f25`.

[58] Marco Scutari. "Bayesian network constraint-based structure learning algorithms: Parallel and optimised implementations in the bnlearn r package". In: *arXiv preprint arXiv:1406.7648* (2014).

[59] James L. Smith. *OPECS Market Role: Changing Signs?* EIA Workshop on Financial & Physical Oil Market Linkages. Southern Methodist University, Dallas Texas. Sept. 19, 2017. URL: `https://www.eia.gov/finance/markets/reports_presentations/2017/yang.pdf` (visited on 04/10/2018).

[60] Federal Reserve Bank of St. Louis. *FRED Economic Data, St. Louis FRED*. 2018. URL: `https://fred.stlouisfed.org` (visited on 04/02/2018).

[61] *Statistical Programs of the United States Government*. United States Office of Management and Budget, 2017. URL: `\url{https://obamawhitehouse.archives.gov/sites/default/files/omb/assets/information_and_regulatory_affairs/statistical-programs-2017.pdf}`.

[62] *Stochastic Processes, Markov Chains,and Markov Models*. 2009. URL: `http://cl.indiana.edu/~md7/09/645/slides/06-markov/06-markov-2x3.pdf`.

[63] Peter Norvig Stuart Russell. *Artificial Intelligence: A Modern Approach*. 1st. Prentice Hall, 1995. ISBN: 0131038052, 9780131038059. URL: `http://gen.lib.rus.ec/book/index.php?md5=BCF9E109BBFA056D2AFDB40BFB87A879`.

[64] Ioannis Tsamardinos, Laura E Brown, and Constantin F Aliferis. "The max-min hill-climbing Bayesian network structure learning algorithm". In: *Machine learning* 65.1 (2006), pp. 31–78.

[65] Ravi Kannan Venkatesan Guruswami. *Lecture Notes: Markov Chains*. 2012. URL: `https://www.cs.cmu.edu/~venkatg/teaching/CStheory-infoage/book-chapter-5.pdf`.

[66] Qiuping Wang, Hao Sun, and Qi Zhang. "A Bayesian Network Model on the Public Bicycle Choice Behavior of Residents: A Case Study of Xian". In: *Mathematical Problems in Engineering* 2017 (2017).

[67] Jian Yang. *The impact of crude oil inventory announcements on prices: evidence from derivatives markets*. EIA Workshop on Financial & Physical Oil Market Linkages. Leeds School of Business, University of Colorado Boulder. Sept. 19, 2017. URL: https://www.eia.gov/finance/markets/reports_presentations/2017/yang.pdf (visited on 04/10/2018).

[68] ChengXiang Zhai. "A Brief Note on the Hidden Markov Models (HMMs)". In: *Department of Computer Science, University of Illinois at Urbana-Champaign, IL, USA* (2003).